\documentclass[%
reprint,
superscriptaddress,
amsmath,amssymb,
aps,
prb,
floatfix,
]{revtex4-2}
\usepackage{silence}
\usepackage{graphicx}
\usepackage{dcolumn}
\usepackage{bm}
\usepackage{xcolor}
\usepackage{amsmath}
\usepackage{dirtytalk}
\usepackage{tabularx}
\usepackage{multirow}
\usepackage{makecell}
\usepackage{float}
\usepackage{placeins}
\usepackage[version=4]{mhchem}
\usepackage{hyperref}

\begin{document}

\preprint{APS/123-QED}

\title[PRB]{Lifetime effects and satellites in the photoelectron spectrum of platinum metal}

\newcommand{\OxfordChem}{Department of Chemistry, University of Oxford, Inorganic Chemistry Laboratory, South Parks Road, Oxford, OX1 3QR, United Kingdom}
\newcommand{\UCLChem}{Department of Chemistry, University College London, 20 Gordon Street, London, WC1H 0AJ, United Kingdom}

\author{P.~Bhatt}
\affiliation{\protect\UCLChem}
\affiliation{\protect\OxfordChem}
\affiliation{Istituto Officina dei Materiali (IOM)-CNR, Laboratorio TASC, in Area Science Park, S.S.14, Km 163.5, Trieste I-34149, Italy}

\author{J.~J.~Gutiérrez Moreno}
\affiliation{Barcelona Supercomputing Center, Plaça Eusebi Güell 1-3, 08034 Barcelona, Spain}

\author{L.~E.~Ratcliff}
\affiliation{Centre for Computational Chemistry, School of Chemistry, University of Bristol, Bristol BS8 1TS, United Kingdom}
\affiliation{Hylleraas Centre for Quantum Molecular Sciences, Department of Chemistry, UiT The Arctic University of Norway, N-9037 Tromsø, Norway}

\author{A.~A.~Riaz}
\affiliation{\protect\UCLChem}

\author{C.~M.~L.~André}
\affiliation{\protect\OxfordChem}
\affiliation{Université Paris-Saclay, 9 Rue Joliot Curie, Gif-sur-Yvette, 91190 France}

\author{A.~S.~Y.~Lu}
\affiliation{\protect\OxfordChem}

\author{R.~G.~Palgrave}
\affiliation{\protect\UCLChem}
\affiliation{HarwellXPS, Research Complex at Harwell, Rutherford Appleton Laboratory, Didcot OX11 0DE, United Kingdom}

\author{A.~Gloskovskii}
\author{C.~Schlueter}
\affiliation{Deutsches Elektronen-Synchrotron (DESY), Notkestraße 85, Hamburg 22607, Germany}

\author{P.~K.~Thakur}
\author{T.-L.~Lee}
\affiliation{Diamond Light Source Ltd., Diamond House, Harwell Science and Innovation Campus, Didcot, OX11 0DE, United Kingdom}

\author{A.~Regoutz}
\affiliation{\protect\UCLChem}
\affiliation{\protect\OxfordChem}

 \email{anna.regoutz@chem.ox.ac.uk}

\date{\today}

\begin{abstract}
This work presents a comprehensive investigation of the electronic structure and many-body photoemission effects in metallic platinum using reflection high-energy electron energy-loss spectroscopy (RHEELS), soft X-ray photoelectron spectroscopy (SXPS), and hard X-ray photoelectron spectroscopy (HAXPES), supported by  \textit{ab initio} calculations. Shallow and deep core state spectra enable the systematic characterisation of intrinsic line-shape asymmetries and satellite structures. Correlation of photoelectron satellites with RHEELS loss features allows the assignment of interband transitions, surface and bulk plasmons, plasmonic overtones, and semi-core ionisation losses across the Pt spectrum. Several previously unresolved satellite features and spin-orbit splittings are identified and discussed. Comparison of experimental valence band spectra with orbital-projected densities of states calculated using \textit{ab initio} density functional theory (DFT) and G\textsubscript{0}W\textsubscript{0} approaches, with and without spin-orbit coupling, demonstrates the critical role of relativistic effects in reproducing the Pt valence electronic structure. Together, these results establish a unified, internally consistent spectroscopic reference for metallic platinum, providing a robust framework for interpreting photoelectron spectra of Pt-containing catalysts, electronic materials, and related 5\textit{d} transition metal systems.
\end{abstract}

\maketitle

\section{\label{sec:Intro}Introduction}
Platinum (Pt) is a versatile and important 5\textit{d} transition metal, known for its corrosion resistance, malleability, and ductility.~\cite{wise1950platinum} It finds extensive use in catalysis, chemical sensing, drug delivery, and microelectronics. In catalysis, Pt bimetallic catalysts, \ce{Pt-M} (M = Co, Ni, Pd, Ag)  and \ce{Pt/Al2O3} are used in hydrocarbon reformation,~\cite{yu2012review} or CO oxidation in catalytic converters relevant for the motor vehicular industry.~\cite{an2014design,he2024single} In sensing, Pt thin films are often used to detect the presence of various gases,~\cite{rong2021pt} while in electronics, electrodes and connectors often use Pt due to its high electrical conductivity.~\cite{johnsson2022review} Pt is also known for its potential in antitumour therapy as a drug delivery compound.~\cite{akhtar2022improvement,mason2023platinum,jiao2023enhancing,jin2024development} Despite its high cost and scarcity, platinum in many cases is the most optimal material for these applications. This necessitates holistic knowledge of its electronic structure, which governs its functionality,~\cite{wach2020comparative} and provides insights into the behaviour of Pt and Pt-based materials in their many applications.\par

To this end, experimental and theoretical approaches have been utilised to elucidate the electronic structure of Pt. Experimentally, X-ray photoelectron spectroscopy (XPS) has been performed on a variety of platinum-based materials. XPS, a surface-sensitive technique, can probe the elemental, chemical, and electronic characteristics of various materials.~\cite{isaacs2021advanced} Often, multi-metallic Pt compounds are measured using XPS pre- and post-catalysis to study the changes in the chemical state of Pt-based compounds.~\cite{yang2023revealing,siburian2021loading} In these studies, the Pt~4\textit{f} orbital is the most common core line measured due to its relative intensity and narrow intrinsic line width, with XPS generally conducted using Mg~K$\alpha$ (1254.6~eV) or Al~K$\alpha$ (1486.7~eV) X-ray energies. In addition, some studies measure the metal as a reference. Whilst this is useful for comparison to compound and mixed-metal data, the presented data often lack complete reporting of all accessible core levels, thereby limiting holistic analysis of Pt itself.~\cite{vovk2017xps} Using multiple core levels can prove beneficial for analysing complex Pt-based materials, but to the best of our knowledge, it remains unreported for XPS conducted with conventional laboratory-based spectrometers.\par

Beyond Mg~K$\alpha$ and Al~K$\alpha$ X-ray energies, some works in the literature have employed hard X-ray photoelectron spectroscopy (HAXPES). In 2009, Anniyev~\textit{et al}.\ reported the valence band (VB) spectrum from a polycrystalline Pt metal foil at 8~keV.~\cite{anniyev2010complementarity} They describe the valence electronic structure comprising broad, partially occupied \textit{d}-states, supported with X-ray absorption near-edge structure (XANES) spectroscopy data. Rumaiz~\textit{et al}.\ reported the Pt~4\textit{f} and VB spectra for a reference Pt thin film using a 2~keV photoexcitation energy.~\cite{rumaiz2023interface} The line shape of Pt~4\textit{f} was described as asymmetric, with a `long high energy tail', corresponding to the shape of metallic core lines.~\cite{zheng2023haxpes,Dig_Pt} The authors state that the partially filled 5\textit{d} state of the VB overlaps with 6\textit{s} states at the valence band maximum, causing complexities in its extrapolation from the Fermi energy, E\textsubscript{F}. Additionally, work by Ueda and Hamada~\cite{ueda2022polarization} combined HAXPES and theory of Pt metal (amongst other 5\textit{d} transition metals) to describe the contributions of the 6\textit{s}, 6\textit{p}, and 5\textit{d} states to the VB. They used polarisation-dependent HAXPES and density functional theory (DFT) to achieve this. While these HAXPES studies have been used to elucidate the valence electronic structure, they often lack a complete description of all accessible core lines, thereby limiting understanding of the electronic structure of Pt. A major advantage of HAXPES is the ability to access deep core lines. While the aforementioned HAXPES literature has not reported the deeper core levels for Pt,  two other works have.~\cite{smith1974photoemission,zborowski2022reference} The deep core levels Pt~3\textit{d}, 3\textit{p}, and 3\textit{s} were collected on as-sputtered Pt films. To the best of our knowledge, these references are the only known sources that report the deep core levels of Pt. However, neither describes the core line shapes and features of the valence state spectra in combination. Despite the extensive experimental efforts for the collection of photoelectron spectra at different X-ray energies, there are still aspects of the spectra of Pt and its electronic structure that are not completely understood.\par 

Mainly, the sparse exploration of core level and valence states means the literature misses out on the discussion of satellites observed in core level spectra. Satellites arise principally from final-state effects caused by energy-loss mechanisms such as plasmon excitation and/or interband transitions. They often appear in the higher-binding-energy (BE) tail of the main photoionisation peak(s). Satellites often pose challenges for XPS analysis if they are not accounted for. Work on electron energy-loss spectroscopy (EELS) of Pt previously reported the presence of energy-loss features.~\cite{rudberg1930characteristic,lynch1968characteristic} However, the EELS features have never been correlated to satellites seen in photoelectron spectra. The absence of detailed descriptions of the photoelectron spectra and associated satellites for Pt metal across different X-ray energies provides a strong case for revisiting and updating the experimental literature for platinum.\par 

In addition to experimental spectroscopic methods, the electronic structure of Pt has been explored using theoretical approaches. The use of theoretical \textit{ab initio} calculations expands the understanding of the electronic structure when compared against an experimental VB spectrum. The work of Ueda and Hamada described previously approached the understanding of the valence electronic structure by comparison of theory and experiment. Theoretical calculations often include the band structure, Fermi surface, and density of states (DOS) for Pt. Some literature has specifically endeavoured to understand the band structures of the 5\textit{d} metals, such as works by Papaconstantopoulos,~\cite{papaconstantopoulos1986handbook}  Mueller~\textit{et al}.,~\cite{mueller1971quadratic} and Smith.~\cite{smith1974photoemission} They showed that \textit{d}-states dominate the VB, particularly noting the hybridisation between the 6\textit{s} and 5\textit{d} orbitals near the Fermi edge. In the cases of Mueller~\textit{et al.}\ and Smith, the theoretically calculated DOS of Pt were compared against the experimental spectrum. However, this revealed that the experimental VB at the time (owing to its lower energy resolution) could not capture all VB features. While the pure metal has not been studied in the literature since, theoretical calculations are now reported on Pt-based compounds, such as bimetallic systems,~\cite{wu2020electronic,froidevaux1968electronic} or complexes.~\cite{zhang2022platinum,ienco2023role} These aim to understand the electronic structure of these materials for their application as high-entropy alloys in catalysis or as complexes for drug delivery. As Pt-based compounds are leading materials in the aforementioned fields, a detailed and up-to-date characterisation and description of the Pt metal photoelectron spectra, including satellite structure, multiple core levels, and valence band spectra, obtained by XPS, can serve as a reliable reference for these communities. In particular, comparing the satellite and electronic structures with theoretical calculations and other energy-loss spectroscopic methods can elucidate the mechanisms underlying various transitions in the photoemission process for Pt metal. All this will aid in better understanding the spectra for Pt and Pt-containing compounds. Conducting XPS at various photon energies enables broader observations, based on differences in the contributions of electronic states arising from changes in the photoionisation cross-sections ($\sigma_i$).

This paper presents a comprehensive investigation of the electronic structure of metallic Pt using electron energy-loss spectroscopy (EELS) and photoelectron spectroscopy (PES) to establish a reliable, internally consistent reference dataset. Building on a prior study of W metal, which demonstrated the importance of carefully disentangling satellite features and valence band structure in metallic XPS spectra,~\cite{kalha2022lifetime} we apply a similar multi-technique approach to Pt. By combining soft and hard X-ray photoelectron spectroscopy, we systematically probe semi-core and core levels, enabling a unified description of line shapes, satellite structures, and their energy-loss origins. The valence band is analysed in detail through comparison with orbital-weighted electronic densities of states calculated using multiple \textit{ab initio} approaches, including DFT and G\textsubscript{0}W\textsubscript{0} methods. Correlating core-level satellites with EELS-derived loss features and theoretical electronic structure provides new insight into final-state effects and interband transitions in Pt. Together, these results provide a comprehensive, critically assessed spectroscopic reference for metallic platinum across multiple photon energies, supporting a more rigorous interpretation of Pt and Pt-based materials in catalysis, electronics, and related applications.\par

\section{Methodology}
\subsection{Experimental Methodology} 
A polycrystalline platinum foil (99.9~\% metal basis, 0.025~mm thick, Thermo Scientific, UK) was used for all experiments. It was mounted using conductive carbon tape for all experiments, and the surface was sputter-cleaned until the C and O signals were minimised.

Reflection high-energy electron energy loss spectroscopy (RHEELS) was conducted at beamline P22 at PETRAIII, German Electron Synchrotron DESY in Hamburg, Germany.~\cite{schlueter2019new} The sample was argon sputter cleaned with a focused Ar\textsuperscript{+} ion gun operating with a 3~keV voltage, rastering over a 2~$\times$~2~mm\textsuperscript{2} area. A high-energy electron gun (Kimball EMG-4212) delivering an electron beam at 6~keV with a spot size of $\approx$100~$\mu$m$^2$ was used. The primary electron energy spread of the \ce{LaB6} cathode was $\approx$0.5~eV. A Phoibos 225HV analyser (SPECS, Berlin, Germany) was used with the small area lens mode and a slit size of 3~mm. The total energy resolution at 6.0~keV in this setup was determined to be 599~meV, using the full width half maximum (FWHM) of the main elastic peak. 16/84\% international standard method of resolution determination with a polycrystalline gold (Au) foil.~\cite{wolstenholme2008summary} This is seen in Figure~\href{SI.pdf#fig:Au}{S1}(a) of the Supplementary Information. \par

SXPS and HAXPES were conducted at beamline I09 at Diamond Light Source, UK.~\cite{Duncan_2018} The Pt foil was sputtered \textit{in-situ} using a defocused Ar\textsuperscript{+} ion gun, operating with 1.5~keV acceleration voltage and an emission current of 10~mA. SXPS and HAXPES measurements employed photon energies of 1.700~keV and 5.927~keV, respectively. These are hereafter referred to as 1.7~keV and 5.9~keV for simplicity. 1.7~keV was selected using a plane-grating monochromator with 400~lines/mm, and 5.9~keV was selected using a double crystal Si~111 monochromator in conjunction with a Si~004 channel-cut post-monochromator, providing total experimental resolutions of 355$\pm$10~meV and 258$\pm$10~meV, respectively. This is seen in Figure~\href{SI.pdf#fig:Au}{S1}(b-c) of the Supplementary Information. The end station is equipped with a VG Scienta EW4000 electron analyser with a $\pm$28$^{\circ}$ angular acceptance. All measurements were performed at incident photon angles of 15$^\circ$ and near-normal emission. 

Pass energies of 50 and 200~eV were used for SXPS and HAXPES experiments, respectively. The BE scale was aligned to the intrinsic Fermi energy (E\textsubscript{F}) of the Pt sample. The probing depths at 1.7~keV and 5.9~keV energies were estimated based on theoretical total inelastic mean free path (IMFP, $\lambda$) values using the non-relativistic TPP-2M formula as implemented in QUASES-IMFP-TPP2M~v3.0.~\cite{tanuma1994calculations} The obtained $\lambda$ values were 1.77 and 4.67~nm for 1.7~keV and 5.9~keV, respectively, giving probing depths $3\lambda$ of 5.3 and 14.0~nm. The input parameters for the QUASES software are summarised in Table~\href{SI.pdf#table:Quases}{S1} of the Supplementary Information.

 \subsection{Computational Methodology}

To explore the electronic structure of Pt metal, theoretical approaches were combined with experimental results. All \textit{ab initio} calculations were carried out using the QuantumATK package.~\cite{smidstrup2019quantumatk} The bulk Pt system was modelled as a face-centred cubic (FCC) structure with a lattice constant of 3.92~\AA . DFT calculations employed the \textit{high}-accuracy linear combination of atomic orbitals (LCAO) basis set for Pt (18 basis functions per atom), paired with the PseudoDojo norm-conserving pseudopotential.~\cite{van2018pseudodojo} This combination is optimised to provide a good balance between computational efficiency and high precision. The generalised gradient approximation (GGA) and the Perdew-Burke-Ernzerhof for Solids (PBEsol)\cite{perdew2008PBEsol} exchange-correlation functionals were used. The density mesh cut-off was set to 125~Ha (3401.42~eV) with a k-point grid of 20$\times$20$\times$20.A highly dense k-point grid and energy cut-off were employed to ensure the convergence of the G\textsubscript{0}W\textsubscript{0} self-energy and accurately capture the fine structure of the Pt electronic states. The electronic smearing was handled using the Fermi-Dirac occupation method with a broadening of 300~K. The electronic structure was investigated by calculating the projected density of states (PDOS). Two different approaches were used to calculate the PDOS, DFT and G\textsubscript{0}W\textsubscript{0} approximations. Both DFT and G\textsubscript{0}W\textsubscript{0} calculations were also performed with spin-orbit coupling (SOC) included, as it was found to be particularly relevant for accurately describing the electronic properties of Pt due to its high atomic mass. Single-shot G\textsubscript{0}W\textsubscript{0} calculations were performed,  using the DFT ground state as the starting point... An energy cut-off of 50~eV was used for the G\textsubscript{0}W\textsubscript{0} calculations, with the same k-point sampling. The unweighted PDOS is provided in the Supplementary Information~\href{SI.pdf#fig:DOS}{S2}.

\subsection{\label{X_section_method}Comparison of Theory and Experiment}

To directly compare the experimentally measured VB spectra and the theoretically calculated PDOS, both were aligned to the experimental Fermi edge (E\textsubscript{F} = 0~eV) of Pt from PES. Additionally, the individual orbital contributions in the PDOS were broadened and weighted to match the experimental data. For all approaches, the theoretical PDOS spectra are first broadened with a Gaussian function to match the total energy resolution of the experiment, as reported using the 16/84 method above, which was applied in the Galore software package.~\cite{j2018galore} Then, the orbital contributions of the PDOS were photoionisation cross-section ($\sigma_{i}$) weighted using four different approaches, following previous work by the authors on W metal.~\cite{kalha2022lifetime} 

The limitation motivating the exploration of multiple approaches is that these cross-sections are available only for states occupied in the atom's ground state. The highest occupied orbitals in Pt for which cross-sections are available are therefore the 5\textit{p}, 5\textit{d}, and 6\textit{s} states. Whilst the 5\textit{d} and 6\textit{s} states do indeed contribute to the valence states, the 5\textit{p} states constitute a shallow core level at BEs of 66.9 and 51.3~eV, unlikely to play a role in the VB. Instead, the \textit{p}-state contributions arise from unoccupied states of 6\textit{p} character, for which cross-sections are not available. Therefore, alternative approaches were employed, referred to as (1) Scofield, (2) Pb correction, (3) least-mean square (LMS), and (4) Ueda. Approach (1) uses the photoionisation cross-sections of the occupied Pt~5\textit{p}, 5\textit{d}, and 6\textit{s} orbitals as calculated by Scofield without any further adjustment.~\cite{Scofield1973,j2018galore,Kalha2020} Approach (2) involves using the ratio of $\sigma_{i}$ values from Pb~6\textit{p}/Pb~6\textit{s} to estimate the Pt~6\textit{p} cross-section relative to the known Pt~6\textit{s} one. Pb is chosen because it is the next element with both of these orbitals occupied. Approach (3) uses the Scofield $\sigma_{i}$ values for 6\textit{s} and 5\textit{d} and applies a least-mean square optimisation between the experimental and simulated spectra to arrive at an appropriate value for the 6\textit{p} cross-section. Lastly, approach (4) uses the relative, optimised photoionisation cross-sections determined by Ueda~\textit{et al.}~\cite{ueda2022polarization} The so corrected PDOS are then summed to provide an overall, $\sigma_{i}$-corrected PDOS which can be compared to the spectral envelope of the experimental VBs.

\section{Results and Discussion}
\subsection{\label{sec:REELS}Reflection High Energy Electron Energy
Loss Spectroscopy: RHEELS}

\begin{figure}[htbp]
\centering
    \includegraphics[keepaspectratio, width = \linewidth]{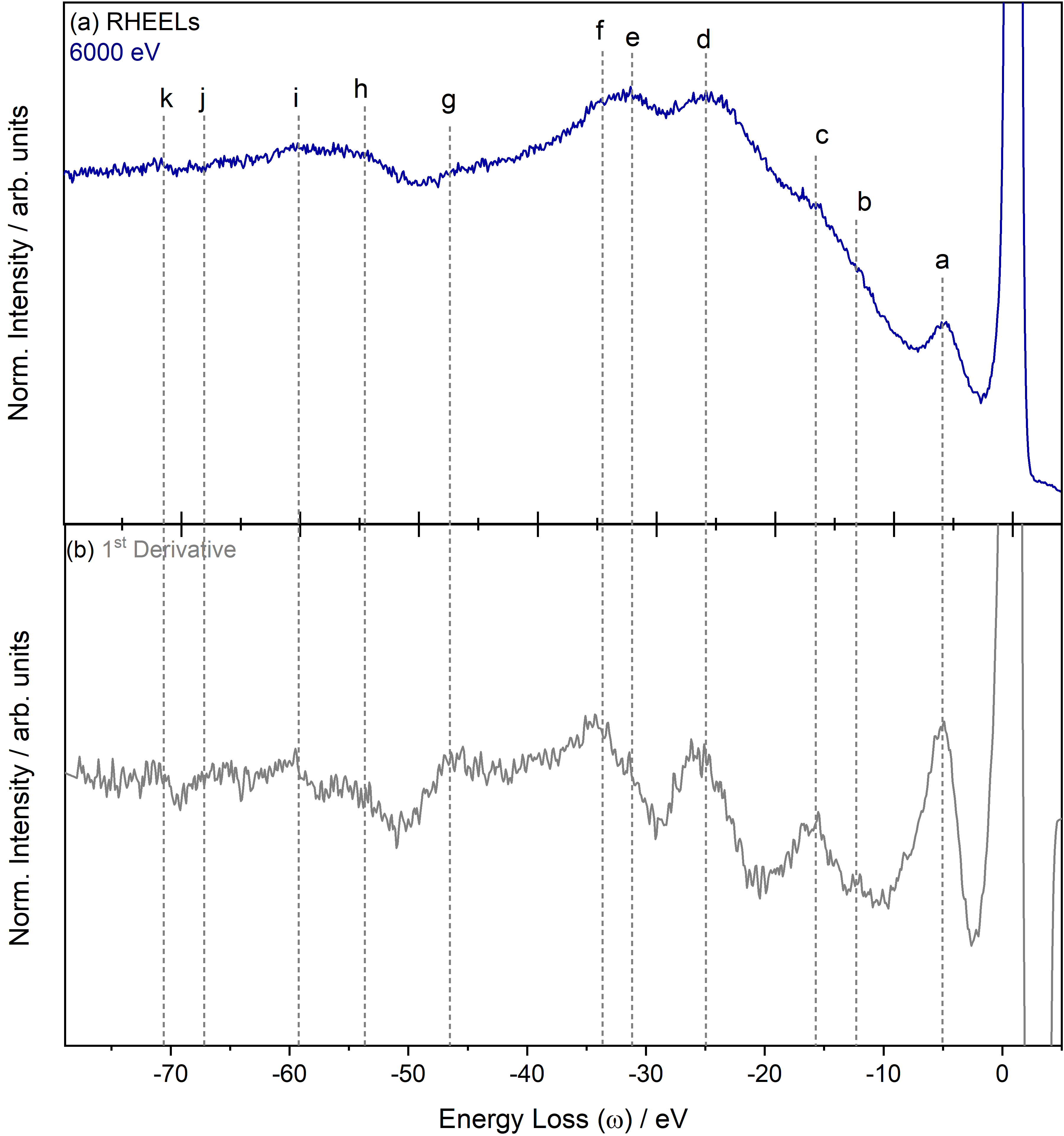}
    \caption{RHEELS of metallic platinum. (a) Collected RHEELS spectrum, and (b) its first derivative with respect to energy loss. The energy-loss features are labelled \textbf{a}-\textbf{k}. The spectrum is aligned to the primary elastic peak. The first-derivative curve was smoothed using a 2\textsuperscript{nd}- order Savitzky-Golay method over a window of 22 data points.}
    \label{fig:RHEELS}
\end{figure}

To guide the interpretation of the satellite features observed in the collected X-ray photoelectron spectra, reflection high-energy electron energy loss spectroscopy (RHEELS) was acquired for Pt. Figure~\ref{fig:RHEELS}(a) displays the collected RHEELS spectrum, marked with its main energy loss peaks (or features) labelled \textbf{a}-\textbf{k}. Their values and assignments are tabulated in Table~\ref{tab:RHEELS_peak}. Figure~\ref{fig:RHEELS}(b) shows the first derivative of the RHEELS spectrum to aid in the determination of the energy loss positions (\textit{w}) of the observed features. 

Whilst both low and high energy EELS were collected on the Pt sample, the RHEELS spectrum is chosen for the main discussion because its structures are better resolved than those of REELS measured at lower electron energies. For example, features \textbf{h} and \textbf{i} can be seen more easily here (see Figures~\href{SI.pdf#fig:REELsoverview}{S3} and \href{SI.pdf#fig:REELs}{S4} of the Supplementary Information for the soft X-ray REELS data). This is attributed to differences in the background, with the low energy spectra exhibiting a higher background due to increased inelastic scattering. In any case, the observable features in the low energy REELS data are in agreement with the RHEELS discussed here.\par

\begin{table}[ht]
    \caption{\label{tab:RHEELS_peak}Energy loss (\textit{w}) peak positions and assignments of loss features of Pt metal determined from RHEELS in Figure~\ref{fig:RHEELS}. The estimated error for the reported energy positions is $\pm$0.3~eV for the low-loss features. The semi-core states are subject to greater uncertainty due to considerable broadening.}
    \begin{ruledtabular}
    \resizebox{\columnwidth}{!}{%
    \begin{tabular}{ccc}
    \textbf{Feature} & \textbf{\textit{w}~/~-eV} & \textbf{Assignment}  \\
    \hline
    \textbf{a} & 4.8 & Interband transition \\
    \textbf{b} & 13.4 & Interband transition \\
    \textbf{c} & 16.4 & Interband transition \\
    \textbf{d} & 24.5 & Surface plasmon \\
    \textbf{e} & 31.0 & Interband transition\\
    \textbf{f} & 34.8 & Bulk plasmon\\
    \textbf{g} & 46.7 & Plasmonic overtone\\
    \textbf{h} & 53.4 &  5\textit{p}\textsubscript{3/2} ionisation\\
    \textbf{i} & 59.5 & Plasmonic overtone \\
    \textbf{j} & 68.2 &  5\textit{p}\textsubscript{1/2} ionisation\\
    \textbf{k} & 71.3 &  4\textit{f}\textsubscript{7/2} ionisation\\
    \end{tabular}
    }
    \end{ruledtabular}
\end{table}

In order to provide context and comparison to the literature on Pt loss features, an outline of the existing literature is highlighted first, with \textit{w} values summarised in Table~\href{SI.pdf#tab:reelslit}{S2} of the Supplementary Information. It should be noted that the energy loss position from the main elastic loss peak for the surface and bulk plasmons of Pt are 24.1 and 34.1~eV, respectively, based on the theoretical plasmon model introduced by Pines in Ref.~\cite{pines1956collective}. These assume a `free electron' configuration of Pt with 10 valence electrons (5\textit{d}\textsuperscript{9}6\textit{s}\textsuperscript{1}).\par

The first reported EELs on Pt was conducted by Rudberg in 1930,~\cite{rudberg1930characteristic} reporting loss features 6.6, 9.4, 11.7, 24.8, 33.7, and 34.8~eV from the main elastic loss peak. Similarly, Möllenstedt measured characteristic energy losses at 14, 22.4, 46 and 61.4~eV.~\cite{mollenstedt1949electrostatic,albert1956einfluss} Between these earliest recorded works, there is a marked difference in the observation of loss feature positions in the EELs spectrum. Differences in the positions of the energy loss peaks could be attributed to the preparation of samples, as Rudberg's sample required etching away silver from a Pt substrate to measure Pt.\par 

Work in the 1950s by Gauthé and Kleinn also reported values of some Pt loss features.~\cite{gauthe1958contribution,kleinn1954energiespektren} Gauthé reported values of 18.5, 23.9, 30, 37, and 47~eV, while Kleinn only states the presence of two features at 5.2 and 22.6~eV. Additionally, Powell noted features that partially align with results from both Gauthé and Klein at 6.2, 14.3, and 22.4~eV.~\cite{powell1960origin}\par

The work by Seignac and Robin was the first to investigate the origins of observed loss features.~\cite{seignac1972proprietes} Peaks at 3.6, 6.3, 9.8, 14.5, and 19.8~eV were associated with interband transitions. The feature at 24~eV was interpreted as a surface plasmon loss, and one at 33.5~eV, as a bulk (or volume) plasmon. Similarly, Lynch and Swan attribute peaks at 7.8~eV and 28.2~eV to interband transitions and bulk plasmon loss, respectively.~\cite{lynch1968characteristic} A feature at 55.2 was determined to be due to the ionisation of the O\textsubscript{III}-level (5\textit{p}\textsubscript{3/2}). Lastly, a feature reported at 73.4~eV was tentatively assigned to `G' peaks arising from transitions in excited states.\par 

A key investigation by Schröder~\textit{et al}.\ collected EELs on Pt (111) concluded on the presence of seven energy loss features in their spectrum, at 7.4, 13.5, 24.8, 31.8, 45.1, 54.1, and 71.2~eV.~\cite{schroder1974investigation} These values were compared against some of the aforementioned literature. Schröder and colleagues suggested that the feature at 7.4~eV arises from an interband transition, which is consistent with work by Seignac and Robin. Features at 24.8 and 31.8~eV were ascribed to surface and bulk plasmon losses, respectively. Two features, at higher energy loss positions of 54.1 and 71.2~eV, were tentatively suggested to arise from optical transitions that involve excitations to the O\textsubscript{III} (5\textit{p}\textsubscript{3/2}), N-\textsubscript{VII}, and N-\textsubscript{VI} levels (4\textit{f}), respectively. The literature discussions so far highlight a clear need to develop a holistic understanding of the characteristic energy losses in Pt metal. The vast differences in the reported loss structures indicate that Pt metal might have several features close in energy that are difficult to report clearly.\par

In this work, energy-loss features observed in the RHEELS spectrum indicate additional structures compared to reports in the literature to date. As the features seen in the RHEELS spectrum are much lower in intensity compared to the main elastic electron loss peak in Figure~\ref{fig:RHEELS}, the positions of loss features can be difficult to determine. This is likely why previous literature shows large variation in reported values for loss features. In this work, plotting the first derivative reveals the smaller intensity features more clearly. An example of this is observed for features \textbf{b} and \textbf{c}, which are obscured by the onset of the inelastic loss background in Figure~\ref{fig:RHEELS}(a), but are clearly seen in Figure~\ref{fig:RHEELS}(b) at 13.4 and 16.4~eV, respectively.Additionally, comparing the \textit{w} peak positions between RHEELS collected at 6000~eV electron energy to REELs measured between 250-1000~eV, the positions of the different features in Table~\ref{tab:RHEELS_peak} are consistent for the same polycrystalline Pt metal foil across different energies and spectrometers. Thus, the discussion below of the different features of Pt reported here is compared with the literature to justify the assignments of energy-loss peaks.\par 

Features \textbf{a}, \textbf{b}, and \textbf{c} reported at \textit{w} values of 4.8~eV, 13.4~eV, and 16.4~eV agree well with the interband transitions from Seignac and Robin.~\cite{seignac1972proprietes} The features result from indirect transitions between the \textit{d}-states near the Fermi edge. Corresponding to the earlier theoretical prediction of a surface plasmon at 24.1~eV, RHEELS shows an agreeable feature \textbf{d} present at 24.5~eV. This is additionally supported by experimental reports of the surface plasmon peak between 23-25~eV in the literature described above. Concerning the bulk plasmon, Schröder~\textit{et al}.\ previously described a peak at 31.8~eV as a bulk plasmon loss,~\cite{schroder1974investigation} which was vastly different to the calculated bulk plasmon loss of 34.1~eV. This discrepancy between the experimental and calculated bulk plasmon energy was attributed by the authors to the presence of an additional interband transition within this energy range. From the RHEEL spectrum in Figure~\ref{fig:RHEELS}, two features \textbf{e} and \textbf{f} are observed at \textit{w} = 31.0~eV and 34.8~eV, respectively. To determine which feature is the bulk plasmon, the plots of R(H)EELs collected at 0.5, 1 and 6~keV are utilised (Figure~\href{SI.pdf#fig:REELSzoom}{S5} in the Supplementary Information). Figure~\href{SI.pdf#fig:REELSzoom}{S5} shows an increase in the intensity of feature \textbf{f} when the electron energy increases, while \textbf{e} remains unchanged. Compared to the relative differences in intensity of other interband transitions (\textbf{a},\textbf{b}, and \textbf{c}) and the work by Schröder~\textit{et al.}, where the relative intensity of the bulk plasmon increases with increasing electron energy, feature \textit{e} can be described as this additional interband transition previously deliberated.\par

So far, features \textit{a}-\textit{f} coincide well with the literature. Features with \textit{w} $>35$~eV are mainly reported using X-ray absorption measurements.~\cite{jaegle1969experimental,haensel1969optical} Feature \textbf{g}, seen in RHEELS at 46.7~eV is comparable to features in the works of Schröder~\textit{et al}.\ (45.1~eV) and Möllenstedt (46~eV).~\cite{schroder1974investigation,mollenstedt1949electrostatic} The former noted that the interpretation of this feature was difficult, given its low intensity. This feature was also seen in the isochromats of Pt by Albert,~\cite{albert1956einfluss} and Leder~\textit{et al}.~\cite{leder1956comparison} noting the observation of a feature at $\approx47.3$~eV. These values were extrapolated from characteristic loss features of Au, which shares a similar electronic structure with Pt. For Au, this is assigned to plasmonic overtones from the surface and bulk plasmon losses.~\cite{tehuacanero2016low} Here, this is reported as the same for Pt; since the \textit{w} position is double that of the surface plasmon (\textbf{d}), it is likely an overtone of this feature. Correspondingly, the overtone originating from both the surface and bulk plasmons (\textbf{f}) is likely feature \textbf{i} at 59.5~eV. The observation of plasmonic overtones resulting from multiples and summations of the surface and/or bulk plasmons is noted in Si.~\cite{melinon1997nanostructured} These are also evidenced by the reducing intensity of the overtones at higher \textit{w} seen in Si and here in Pt. These features are only visible in RHEELS (see Figure~\href{SI.pdf#fig:REELs}{S4} in the Supplementary Information that shows the absence of this feature in REELS), likely due to a lower contribution of secondary electrons in the background of the RHEEL spectrum.  Finally, Features \textbf{h}, \textbf{j}, and \textbf{k} match the core ionisation energies of Pt~5\textit{p}\textsubscript{3/2}, Pt~5\textit{p}\textsubscript{1/2} and Pt~4\textit{f}\textsubscript{7/2}, respectively. It is noted that the $\omega$ values from features \textbf{h}, \textbf{j}, and \textbf{k} differ from the reported binding energy positions from HAXPES, with the differences being 2.5, 1.3, and 0.7~eV, respectively. The differences between these values from RHEELS and HAXPES can be understood by considering the interaction of (photo)electrons emitted in both methods. In RHEELS, incident electrons undergo inelastic scattering due to interactions with a material. There is no emission of electrons from the material itself. Whereas, in HAXPES, an ejected photoelectron undergoes additional interaction with its corresponding core hole. This results in a final state effect where the shift of the core level shows hole screening.~\cite{melinon1997nanostructured} A reduction in the differences between the binding energy and the positions for the features from RHEELS going from Pt~5\textit{p}\textsubscript{3/2} to Pt~4\textit{f}\textsubscript{7/2} also indicates that the screening effect decreases going from the semi-core to the localised core states.\par 

With the assignment of all features observed in RHEELS, this work presents the first complete description of the characteristic energy loss features from Pt metal. All features in the RHEEL spectrum have been compared to the various literature sources and assigned as appropriate. This includes a comparative critique of the literature and EELS measurements at different incident electron energies in this work. These established assignments will aid discussions of the satellite features observed in the HAXPES and SXPS of the metal foil, as discussed in the following Sections. 
\subsection{\label{XPS}Core Level Photoelectron Spectroscopy}

The survey spectra displayed in the Supplementary Information (Figures~\href{SI.pdf#fig:Survey}{S6} and~\href{SI.pdf#fig:Survey_hx}{S7}) show that a clean Pt surface was obtained by employment of the Ar\textsuperscript{+} sputtering, with HAXPES showing no discernible O and C signals. In SXPS, apart from the expected presence of Pt, minor contributions from C, O, and Cu were also detected. The supplier, Thermo Scientific, has noted and identified a Cu impurity in the Pt metal at 7~ppm, exceeding the expected Cu trace of 1~ppm.  The low relative intensity of this impurity does not adversely affect the Pt spectra. Table~\ref{tab:BE_pos} lists the BE positions of all measured Pt core levels, the relative positions of satellites from the main quasiparticle peak (Sep.), and the spin-orbit splitting ($\Delta_\textrm{SOS}$) taken from the HAXPES collected spectra. Additionally, the FWHM of the core levels is tabulated. The following discussion of the PES data is organised as follows. First, the BE positions, line shapes, and satellite features from SXPS and HAXPES for the semi-core levels, n = 4 and n = 5, where n is the principal quantum number, will be reported. These core levels and their features will be discussed and reviewed in relation to the available literature. Then, the deeper, n = 3, core levels and their satellite features, accessible only by HAXPES, will be presented. Lastly, a systematic analysis of the line widths from SXPS and HAXPES is presented.\par 

\begin{table}[!htbp]
    \caption{\label{tab:BE_pos}Absolute binding energy (BE) positions of platinum core level (CL) and \textit{observable} satellite (Sat.), the BE separation (Sep.) of the satellite from the main core level, the spin-orbit splitting ($\Delta_\textrm{SOS}$), and FWHM of core level peaks as determined from HAXPES ($h\nu$ = 5.9~keV). Satellite BE positions have an estimated error of $\pm$0.5~eV, while CL positions have an estimated error of $\pm$0.1~eV based on differences in peak width and relative intensity.}
    \begin{ruledtabular}
    \resizebox{\columnwidth}{!}{%
    \begin{tabular}{ccccc}
    CL/Sat. & BE~/~eV & Sep.~/~eV &$\Delta_\textrm{SOS}$~/~eV & FWHM~/~eV \\
    \hline
\textbf{Pt~5\textit{p}\textsubscript{3/2}} & 51.3 & - & - &2.8\\
\textbf{Pt~5\textit{p}\textsubscript{1/2}} & 66.9 & - & 15.6 & 5.8\\
    \\
\textbf{Pt~5\textit{s}} & 102.0 & - & - & 6.6\\
  S\textsubscript{1} &118.5& 16.5 & - & - \\
\\
\textbf{Pt~4\textit{f}\textsubscript{7/2}} & 70.6 & - & - & 0.6\\
 S\textsubscript{1} &77.7& 7.1 & - & - \\
\textbf{Pt~4\textit{f}\textsubscript{5/2}} & 74.0 & - & 3.4 & 0.6\\
 S\textsubscript{2} &81.1& 7.1 & - & -  \\ 
\\
\textbf{Pt~4\textit{d}\textsubscript{5/2}} & 314.7 & - & - & 3.6\\
\textbf{Pt~4\textit{d}\textsubscript{3/2}} & 331.7 & - & 17.0 & 4.9\\
S\textsubscript{1} & 346.6 & 14.9 & - & - \\
S\textsubscript{2} & 355.4 & 23.7 & - & - \\
S\textsubscript{3} & 366.9 & 35.2 & - & - \\
S\textsubscript{4} & 376.6 & 44.9 & - & - \\
S\textsubscript{5} & 387.1 & 55.4 & - & - \\
S\textsubscript{6} & 401.8 & 70.1 & - & - \\
\\
\textbf{Pt~4\textit{d}\textsubscript{5/2}} & 314.7 & - & - &3.6\\
S\textsubscript{1} & 346.6 & 14.9 & - & - \\
S\textsubscript{2} & 355.4 & 23.7 & - & - \\
S\textsubscript{5} & 387.1 & 55.4 & - & - \\
\textbf{Pt~4\textit{d}\textsubscript{3/2}} & 331.7 & - & 17.0 & 4.9\\
S\textsubscript{3} & 366.9 & 35.2 & - & - \\
S\textsubscript{4} & 376.6 & 44.9 & - & - \\
S\textsubscript{6} & 401.8 & 70.1 & - & - \\
\\
\textbf{Pt~4\textit{p}\textsubscript{3/2}} & 519.3 & - & - & 5.0 \\
    S\textsubscript{1} & 546.2& 26.9 & - & - \\
        S\textsubscript{2} & 552.9& 33.6 & - & - \\
            S\textsubscript{3} & 576.8& 57.5 & - & - \\
\textbf{Pt~4\textit{p}\textsubscript{1/2}} & 609.4 & - & 90.1 & 6.0 \\
    S\textsubscript{4} & 636.3& 26.9 & - & - \\
        S\textsubscript{5} & 643.0& 33.6 & - & - \\
            S\textsubscript{6} & 666.9& 57.5 &- &  - \\
\\
\textbf{Pt~4\textit{s}} & 725.2 & - & - & 7.1 \\
    S\textsubscript{1} & 757.1 & 31.9 &- & - \\
        S\textsubscript{2} & 781.9 & 56.7 &- & - \\
\\
\textbf{Pt~3\textit{d}\textsubscript{5/2}} & 2121.7 & - & - &2.8\\
S\textsubscript{1} & 2146.7 & 25.0 & - & - \\
    S\textsubscript{2} & 2155.1 & 33.4 &- & - \\
      S\textsubscript{3} & 2182.2 & 60.5 & - &- \\
\textbf{Pt~3\textit{d}\textsubscript{3/2}} & 2201.9 & - &80.2 & 2.8\\
S\textsubscript{4} & 2226.9 & 25.0 & - & - \\
    S\textsubscript{5} & 2235.3 & 33.4 & - & - \\
        S\textsubscript{6} & 2262.4 & 60.5 & - & - \\
        \\
\textbf{Pt~3\textit{p}\textsubscript{3/2}} & 2646.2 & - &- & 8.2 \\
    S\textsubscript{1} & 2678.0 & 31.8 &- & - \\
        S\textsubscript{2} & 2706.0 & 59.8 & - &- \\
\textbf{Pt~3\textit{p}\textsubscript{1/2}} & 3026.4 & - &380.2 & 11.8 \\
\\
\textbf{Pt~3\textit{s}} & 3297.6 & - &- & 13.7 \\
    S\textsubscript{1} & 3330.1 & 32.5 & - &- \\
       S\textsubscript{1} & 352.9 & 55.3 & - &- \\
    \end{tabular}
    }
    \end{ruledtabular}
\end{table}

\subsubsection{Shallow Core States (BE \texorpdfstring{$<$}{<} 1000~eV)}

\begin{figure*}[htbp]
\centering
\includegraphics[keepaspectratio, width = 0.66\linewidth]{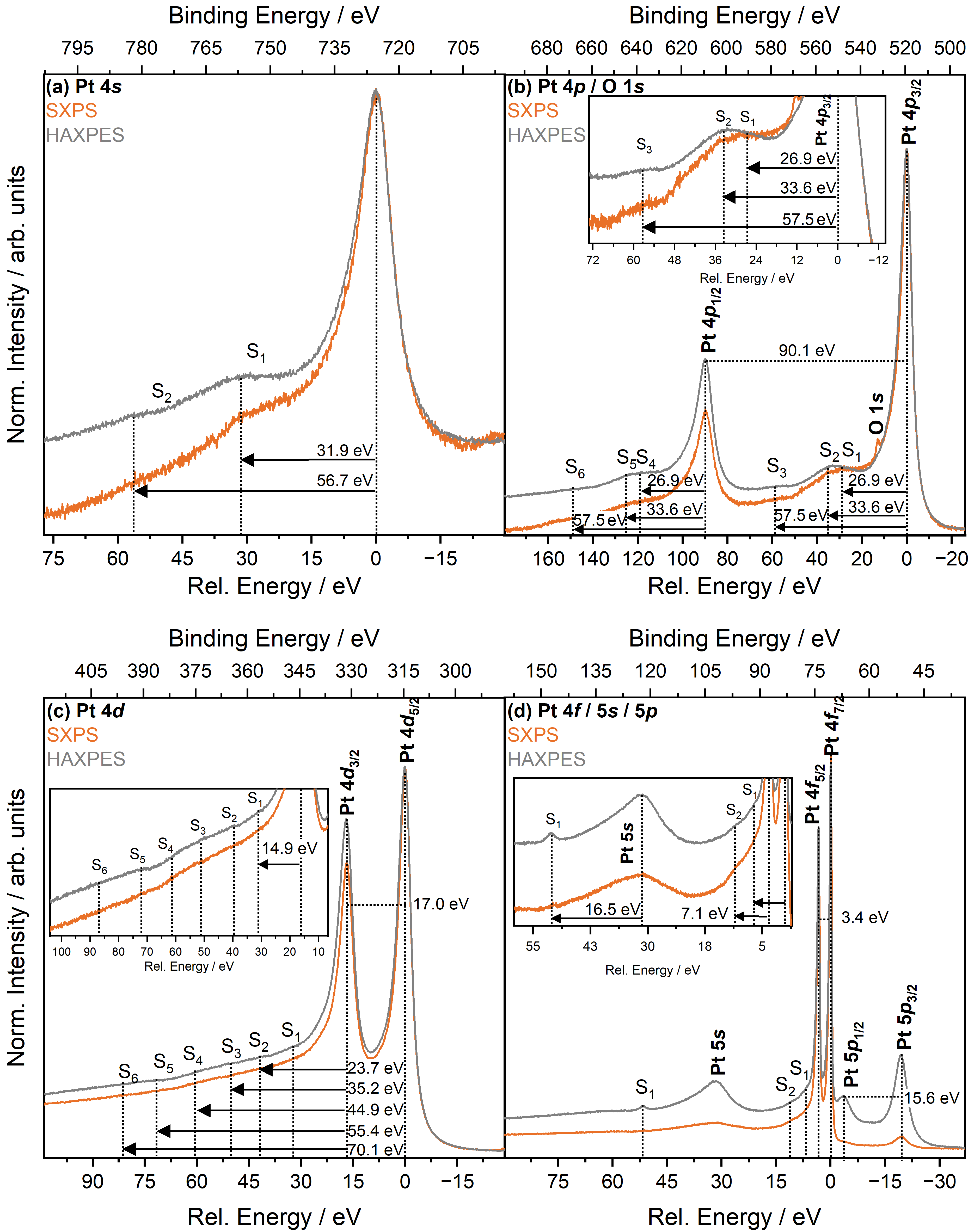}
    \caption{Core level photoelectron spectra including the (a) Pt~4\textit{s}, (b) Pt~4\textit{p} / O~1\textit{s}, (c) Pt~4\textit{d}, and (d) Pt~4\textit{f} / 5\textit{s} / 5\textit{p} synchrotron-based SXPS ($h\nu$ = 1.7~keV) in orange and HAXPES ($h\nu$ = 5.9~keV) in grey from a polycrystalline Pt foil. The spectra are normalised to their maximum height. Spectra are plotted on a relative BE scale, aligning the main photoionisation peak to 0~eV. The experimental BE scale, aligned with the intrinsic E\textsubscript{F} of Pt metal, is displayed above each core-level spectrum. Insets depict expanded regions of the higher-BE tails and satellites, plotted on a logarithmic scale to aid observation of the satellite features discussed in the main text.}
    \label{fig:Ptscl}
\end{figure*}

Although the most frequently accessed core level of Pt in the literature is the Pt~4\textit{f}, other core levels, including Pt~4\textit{s}, 4\textit{p}, 4\textit{d}, 5\textit{s}, and 5\textit{p}, can also be easily measured with laboratory Mg and Al K$\alpha$ X-ray sources as their BEs are well below 1000~eV. These core levels were measured here using SXPS and HAXPES, and are depicted in Figure~\ref{fig:Ptscl}. The asymmetric line shape of core levels from metallic compounds is evident in each core level as a higher BE tail. This line shape is attributed to the interaction between electrons that have undergone shake-up processes after photoejection from the initial core state and the conduction band of metals.~\cite{doniach1970many,hufner1975core,hufner1975xps} All core levels in Figure~\ref{fig:Ptscl} also present a series of satellite features at higher BE. These have been labelled S\textsubscript{N}, where N = 1, 2, 3, etc. The previously drawn conclusions about electron energy losses from RHEELS can be used to interpret some of the structures observed in these core levels. To the best of our knowledge, none of the satellite features presented for these core levels of Pt have been previously reported.\par

The Pt~4\textit{s} core level, shown in Figure~\ref{fig:Ptscl}(a) appears at a BE value of 725.2~eV, in good agreement with past XPS measurements.~\cite{nyholm1980core,karlsson1967electron,zborowski2022reference,zheng2023haxpes,rincon2023platinum} The line width of 7.1~eV derived from the HAXPES spectrum is lower than the FWHM value of 9.8~eV measured by Zheng~\textit{et al}, attributed to improved signal-to-noise ratio of the core level reported in this work.~\cite{zheng2023haxpes} There are two prominent satellites, S\textsubscript{1} and S\textsubscript{2}, present at separations of 31.9~eV and 56.7~eV from the main spectral peak, respectively. Compared to the values of energy loss peaks in Table~\ref{tab:RHEELS_peak}, S\textsubscript{1} aligns well with the position of the interband transition \textbf{e} (\textit{w} = 31.0~eV) / bulk plasmon feature \textbf{f} at (\textit{w} = 34.8~eV) overlap. This is not further distinguished due to the larger FWHM of S\textsubscript{1} compared to the RHEELs spectrum. S\textsubscript{2} corresponds to the 5\textit{p}\textsubscript{1/2} excitation at \textit{w} = 59.5~eV (feature \textbf{i}).\par 

Similarly, the Pt~4\textit{p} core level (Figure~\ref{fig:Ptscl}(b)) presents BE positions of 519.3 and 609.4~eV for 4\textit{p}\textsubscript{3/2} and 4\textit{p}\textsubscript{1/2}, respectively. Again these agree well with previous literature,~\cite{nyholm1980core,karlsson1967electron,zborowski2022reference,zheng2023haxpes,rincon2023platinum} providing a spin-orbit splitting $\Delta_\textrm{SOS}$ value of 90.1~eV. At the higher BE tail of Pt~4\textit{p}\textsubscript{3/2}, the presence of the O~1\textit{s} core level at 533.2~eV is noted in the SXPS spectrum. The signal intensity was minimised by employing Ar\textsuperscript{+} sputtering, as described in the experimental methodology. The sputtering, combined with the deeper probing depth achieved by HAXPES (4.34~nm) compared to SXPS (1.34~nm) for Pt~4\textit{p}\textsubscript{3/2} as calculated by QUASES-IMFP-TPP2M, results in the absence of O~1\textit{s} in the HAXPES spectrum. In both cases, the contribution from O~1\textit{s} is negligible for descriptions of the electronic structure. Pt~4\textit{p} exhibits six satellite structures, S\textsubscript{1}, S\textsubscript{2}, and S\textsubscript{3} corresponding to Pt~4\textit{p}\textsubscript{3/2}, and S\textsubscript{4}, S\textsubscript{5}, and S\textsubscript{6} corresponding to 4\textit{p}\textsubscript{1/2}. These can be understood as two sets of satellites occurring from equivalent events, as the relative BE positions of S\textsubscript{1}, S\textsubscript{2}, and S\textsubscript{3} to its main quasiparticle peak Pt~4\textit{p}\textsubscript{3/2} are equivalent to S\textsubscript{4}, S\textsubscript{5}, and S\textsubscript{6} with Pt~4\textit{p}\textsubscript{1/2}. Thus, descriptions of S\textsubscript{1}, S\textsubscript{2}, and S\textsubscript{3} are sufficient to understand the satellite features arising in the Pt~4\textit{p} spectrum. When the entire Pt~4\textit{p} region is considered in Figure~\ref{fig:Ptscl}(b), it appears that there might only be two satellite features per main core line. However, looking closely at the tail end of Pt~4\textit{p}\textsubscript{3/2} in the inset reveals the close BE positions of S\textsubscript{1} and S\textsubscript{2}. They reside approximately 26.9~eV and 33.6~eV away from the main peak, respectively. The position of these satellites coincides with the energy loss attributed to the surface (\textbf{d}) and bulk plasmons (\textbf{f}) in Section~\ref{sec:REELS}. In comparison, these features are not separated in the Pt~4\textit{s} spectra, where the spectral widths are much greater. The third satellite S\textsubscript{3} at approximately 57.5~eV from Pt~4\textit{p}\textsubscript{3/2} appears broad and low in intensity, corresponding to a plasmonic overtone.\par

The Pt~4\textit{d}\textsubscript{5/2} and Pt~4\textit{d}\textsubscript{3/2} core levels are observed at 314.7 and 331.7~eV, commensurate with other XPS measurements.~\cite{shyu1988identification,resende2011effect,schon1972high,karlsson1967electron,nyholm1980core,schneider1981actinide,resende2011effect} Compared to the n = 4 core levels discussed so far, Pt~4\textit{d} presents a complex set of satellite features at the tail of the 4\textit{d}\textsubscript{3/2} peak. This is due to the much lower $\Delta_\textrm{SOS}$ value of 17.0~eV for Pt~4\textit{d} resulting in overlapping satellite series arising from both 4\textit{d}\textsubscript{5/2} and 4\textit{d}\textsubscript{3/2}. In order to systematically disentangle these satellite features, all relative BE positions reported are referenced with Pt~4\textit{d}\textsubscript{3/2} in Table~\ref{tab:BE_pos}. S\textsubscript{1}, S\textsubscript{2}, S\textsubscript{3}, S\textsubscript{4}, S\textsubscript{5}, and S\textsubscript{6} are reported at BE values of 14.9, 23.7, 35.2, 44.9, 55.4, and 70.1~eV above Pt~4\textit{d}\textsubscript{3/2}. Except for S\textsubscript{1}, the relative BE positions (rel. to Pt~4\textit{d}\textsubscript{3/2}) of all other satellite features coincide with previously described electron loss events for Pt in Section~\ref{sec:REELS}. In ascending order of relative BE position, these features are attributed to the surface plasmon (S\textsubscript{2}), bulk plasmon (S\textsubscript{3}), plasmonic overtones (S\textsubscript{4}), plasmon generation by 5\textit{p} (S\textsubscript{5}) and  4\textit{f} electrons (S\textsubscript{6}). Since an equivalent number of satellite features were observed from each singlet in Pt~4\textit{p}, S\textsubscript{1}-S\textsubscript{6} are likely to be a combination of satellite features from Pt~4\textit{d}\textsubscript{5/2} as well.\par 

Accordingly, Table~\ref{tab:4dref} shows the assignment of all six satellite features identified for Pt~4\textit{d} to each quasiparticle peak. Relative to Pt~4\textit{d}\textsubscript{5/2}, S\textsubscript{1}, (31.9~eV) is associated with the interband transition of Pt. Correspondingly, S\textsubscript{2} (40.7~eV) and S\textsubscript{4} (61.9~eV) coincide with the plasmonic overtones. Additionally, S\textsubscript{3} and S\textsubscript{5} reveal 5\textit{p} and 4\textit{f} ionisation, respectively (when considered from Pt~4\textit{d}\textsubscript{5/2}). Feature S\textsubscript{6} at 87.1~eV is too high in relative energy to belong to any loss feature from the 4\textit{d}\textsubscript{5/2} core level, and can be assigned to events arising from ejected photoelectrons from the 4\textit{d}\textsubscript{3/2} peak. Similarly, S\textsubscript{1} does not arise from Pt~4\textit{d}\textsubscript{3/2} as its position does not correspond to any of the known loss features understood by RHEELS. Identification of loss features for Pt~4\textit{d} employs a combination of RHEELS established previously, with the understanding that doublet core levels show loss features for each quasiparticle peak (as in Pt~4\textit{p}). Here, this permits a systematic and robust assignment of overlapping satellite features.\par

\begin{table*}
    \centering
    \caption{Assignment of satellites S\textsubscript{N} (N = 1-6) observed in the HAXPES photoelectron spectrum of Pt~4\textit{d}. All features are provided with their respective relative binding energy position (Pos.) to each quasiparticle peak, 4\textit{d}\textsubscript{3/2} and 4\textit{d}\textsubscript{5/2}. Where n/a is stated, no features from the RHEEL spectrum indicated the presence of loss events at those relative energy positions.}
    \begin{tabular}{lcccc}
    \hline \hline
      Sat.   & Pos. (rel. 4\textit{d}\textsubscript{3/2}) / eV & Assign. & Pos (rel. 4\textit{d}\textsubscript{5/2}) / eV  & Assign.  \\
      \hline
      S\textsubscript{1}   & 14.9 & n/a & 31.9 &  Interband transition  \\
      S\textsubscript{2}   & 23.7 & Surface plasmon & 40.7 & Plasmonic overtone  \\
      S\textsubscript{3}   & 35.2 & Bulk plasmon & 52.2 &  5\textit{p}\textsubscript{3/2} ion. \\
      S\textsubscript{4}   & 44.9 & Plasmonic overtone & 61.9 &  Plasmonic overtone  \\
      S\textsubscript{5}   & 55.4 & 5\textit{p}\textsubscript{3/2} ion. & 72.4 & 4\textit{f}\textsubscript{7/2} ion.  \\
      S\textsubscript{6}   & 70.1 & 4\textit{f}\textsubscript{7/2} ion.  & 87.1 &  n/a \\
      \hline \hline
    \end{tabular}
    \label{tab:4dref}
\end{table*}

The semi-core levels, Pt~4\textit{f}, 5\textit{s}, and 5\textit{p}, will be discussed together, since they are very close in BE, as seen in Figure~\ref{fig:Ptscl}(d). The decay of photoionisation cross-sections, $\sigma_i\propto E^{-3}$, has previously been used in studies of W metal to aid the interpretation of shallow core levels in SXPS and HAXPES. HAXPES shows an enhancement in the intensities of the 5\textit{s} and 5\textit{p} core levels relative to Pt~4\textit{f}. This corroborates with the ratios of $\frac{\sigma_i(Pt~4\textit{f})}{\sigma_i(Pt~5\textit{s})}$ between  SXPS (19) and HAXPES (11) and $\frac{\sigma_i(Pt~4\textit{f})}{\sigma_i(Pt~5\textit{p})}$ between SXPS (4.3) and HAXPES (2.7).~\cite{Scofield1973} Figure~\href{SI.pdf#fig:cs}{S8} of the Supplementary Information shows $\sigma_i$ values for the core levels discussed, illustrating these relative changes. The improved intensity of the 5\textit{s} and 5\textit{p} core levels in HAXPES enables determination of the BE positions for all n = 5 core levels. In particular, the BE position for Pt~5\textit{p}\textsubscript{1/2} at 66.9~eV with a  $\Delta_\textrm{SOS}$ of 15.6~eV is the first complete report of the Pt~5\textit{p} level. Previous works have only reported the position of the 5\textit{p}\textsubscript{3/2} peak at $\approx51.9$~eV.~\cite{nyholm1980core,zborowski2022reference} This highlights the benefits of conducting SXPS and HAXPES in parallel, making use of different rates of decay in $\sigma_i$ as a function of photon energy.\par

Satellites in the 4\textit{f}/5\textit{s}/5\textit{p} core levels are the least pronounced of the core levels discussed thus far. The inset in Figure~\ref{fig:Ptscl}(d) shows two satellites near the Pt~4\textit{f} and one after the Pt~5\textit{s} core level. The S\textsubscript{1} and S\textsubscript{2} satellites are approximately 7.1~eV away from Pt~4\textit{f}\textsubscript{7/2} and Pt~4\textit{f}\textsubscript{5/2}, respectively. This relative BE position does not directly correspond to an identifiable energy-loss feature observed in RHEELS. However, it does fall within the range of potential interband transitions (between 4.8-16.4~eV).~\cite{seignac1972proprietes} It is not possible to provide a more precise identification of their origin, due to the difficulty in ascertaining their BE position, given their low relative intensity compared to the main photoionisation peak. Additionally, the asymmetric tails of the metallic core levels hinder the observation of distinct satellite features, and overlapping loss features from Pt~5\textit{p} and 4\textit{f} can obscure distinct peaks. Conversely, the S\textsubscript{1} satellite related to Pt~5\textit{s} presents a clearly observable peak that is situated 16.5~eV away, corresponding to the interband transition (feature \textbf{c}) at 16.4~eV observed in RHEELS.\par

\subsubsection{Deep Core Levels}\label{sec:ptdeep}

\begin{figure*}
\centering
    \includegraphics[keepaspectratio, width = \linewidth]{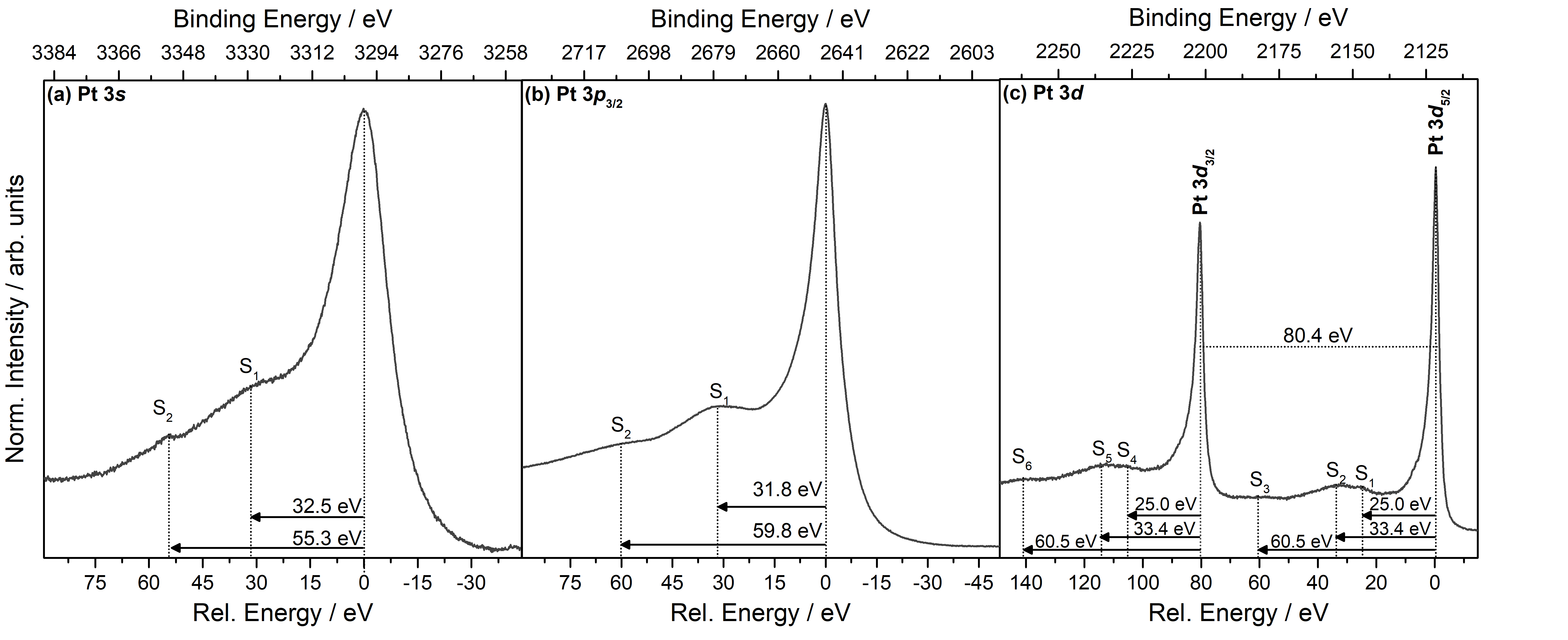}
    \caption{Core level photoelectron spectra, including (a) Pt~3\textit{s}, (b) Pt~3\textit{p}\textsubscript{3/2}, and (c) Pt~3\textit{d} collected from a polycrystalline Pt foil using synchrotron-based HAXPES ($h\nu$ = 5.9~keV). Spectra are plotted on a relative BE scale, aligning the main photoionisation peak to 0~eV. The experimental BE scale aligned to the intrinsic E\textsubscript{F} of the Pt metal is displayed above each core level spectrum.} 
    \label{fig:Pt3l}
\end{figure*}

The Pt 3\textit{s}, 3\textit{p}, and 3\textit{d} core lines can only be accessed by the higher photon energy used in HAXPES. As such, a small number of previous HAXPES studies on Pt have explored them, as described earlier. Figures~\ref{fig:Pt3l}(a), (b), and (c) display the Pt~3\textit{s}, Pt~3\textit{p}\textsubscript{3/2}, and Pt~3\textit{d} HAXPES core level spectra, respectively. As with the shallow core levels discussed previously, metallic Pt exhibits distinct asymmetric peak shapes and no additional Pt chemical states. Thus, the satellite features observed in Figure~\ref{fig:Pt3l} describes the many-electron events in Pt metal resulting from photoexcitation. 

From the BE positions of the main spectral peaks summarised in Table~\ref{tab:BE_pos}, Pt~3\textit{s} is observed at 3297.6~eV (see Figure~\ref{fig:Pt3l}(a)). The work by Zheng~\textit{et al}.~\cite{zheng2023haxpes} reports the same value while Zborowski~\textit{et al}.~\cite{zborowski2022reference} report the BE position at 3298.1~eV. The FWHM by Zheng~\textit{et al}.\ of 13.6~eV agrees well with the value of 13.7~eV observed in this work. Pt~3\textit{p} is observed at BE positions of 2646.2~eV and 3026.4~eV for Pt~3\textit{p}\textsubscript{3/2} (Figure~\ref{fig:Pt3l}(b)) and Pt~3\textit{p}\textsubscript{1/2} (Figure~\href{SI.pdf#fig:Survey_hx}{S6} in the Supplementary Information), respectively. The $\Delta_\textrm{SOS}$ is 380.2~eV and matches literature values.~\cite{zheng2023haxpes} The FWHM of these core levels as reported by Zheng~\textit{et al}. were 8.2 and 11.5~eV for Pt 3~\textit{p}\textsubscript{3/2} and 3~\textit{p}\textsubscript{3/2}. As with Pt~3\textit{s} reported FWHM agree very well with line widths determined in this work, for Pt 3~\textit{p}\textsubscript{3/2} and 3~\textit{p}\textsubscript{3/2} these are 8.2 and 11.8~eV, respectively. Lastly, the Pt~3\textit{d} core level shown in Figure~\ref{fig:Pt3l}(c) reports BE positions of 2121.7~eV and 2201.9~eVfor the 3\textit{d}\textsubscript{5/2} and 3\textit{d}\textsubscript{3/2} lines, respectively. The $\Delta_\textrm{SOS}$ between these is 80.4~eV, agreeing well with previous reports.~\cite{zborowski2022reference,zheng2023haxpes,rincon2023platinum} The FWHM of 2.8~eV for both lines is similar to the literature values of 2.9~eV (Pt~3\textit{d}\textsubscript{3/2}) and 3.0~eV (Pt~3\textit{d}\textsubscript{5/2}).~\cite{zheng2023haxpes}

Concerning the satellite features observed for the deep core levels, Pt~3\textit{s} exhibits two satellite features labelled S\textsubscript{1} and S\textsubscript{2}, located at 32.5~eV and 55.3~eV above the main 3\textit{s} core level, respectively. These correspond to the features in RHEELS at 32.5~eV and 55.3~eV, which were attributed to interband transition / bulk plasmon loss and ionisation from the 5\textit{p}\textsubscript{3/2}, respectively. Additionally, Features in Pt~3\textit{s} correspond to the reported satellites in the Pt~4\textit{s} spectrum above. Similarly, Pt~3\textit{p}\textsubscript{3/2} displays two satellite features S\textsubscript{1} and S\textsubscript{2} 31.8~eV and 59.8~eV above the quasiparticle peak. These electronic loss structures are ascribed to the interband transition / bulk plasmon and the plasmonic overtone from the bulk plasmon, also reported for Pt~4\textit{p} in Figure~\ref{fig:Ptscl}(b). The presence of a surface plasmon is not noted for Pt~3\textit{p}\textsubscript{3/2} as it was seen in  4\textit{p}. The higher kinetic energy of the photoelectron peaks for Pt~4\textit{p} than for 3\textit{p} indicates that the probing depth decreases when deeper core levels are probed. From QUASES-IMFP-TPP2M, the decrease in probing depth is from 4.34~nm to 2.90~nm, thus more scattering events are observable from Pt~4\textit{p} compared to Pt~3\textit{p}.~\cite{van1979bulk} In both Pt~3\textit{s} and Pt~3\textit{p}\textsubscript{3/2}, the most prominent satellite arises from bulk plasmon (S\textsubscript{1}), and appear with similar widths compared to the main quasiparticle peak. In contrast, S\textsubscript{2} appears broader in Pt~3\textit{p}\textsubscript{3/2} than in Pt~3\textit{s}. The wider feature in Pt~3\textit{p} from the plasmonic overtone might be hindering clear observation of the characteristic loss feature from 5~\textit{p}\textsubscript{3/2}. Taking into consideration the probing depth between Pt~3\textit{s} (2.43~nm) and Pt~3\textit{p}\textsubscript{3/2} (2.90~nm), the number of characteristic losses from Pt~3\textit{s} is likely to be less than those from Pt~3\textit{p}. This was also seen for n = 4, where the \textit{s}-orbital was followed by two loss features compared to three from each p-orbital core level. Lastly, for Pt~3\textit{d}, each of the doublet lines shows three comparable satellite features, S\textsubscript{1} \& S\textsubscript{4},  S\textsubscript{2} \& S\textsubscript{5}, and S\textsubscript{3} \& S\textsubscript{6} at 25.0, 33.4, and 60.5~eV from their respective main line. The first set of features at 25.0~eV from the main spectral peak in HAXPES coincides with the RHEELS feature at $\approx23.1$~eV arising from surface plasmon losses. The feature at 33.4~eV was characterised as a bulk plasmon loss feature. Lastly, the feature at 60.1~eV aligns with the plasmonic overtone.

\subsubsection{Evaluation of Core Level Line Widths}

\begin{figure}[htbp]
\centering
\includegraphics[keepaspectratio, width = \linewidth]{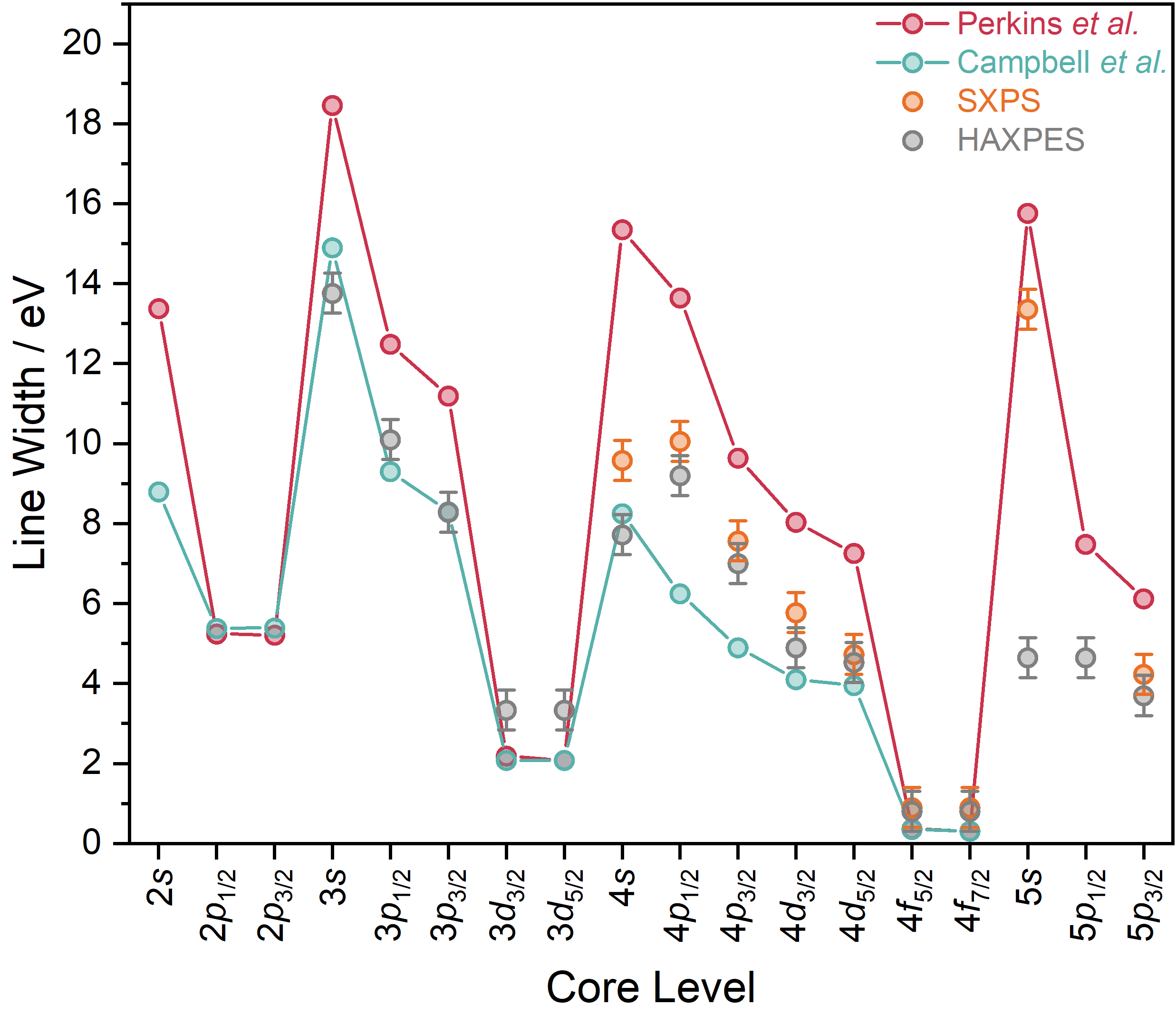}
    \caption{Comparison of the experimental core line width of SXPS ($h\nu$ = 1.7~keV) and HAXPES ($h\nu$ = 5.9~keV) data with reported natural line width values from Refs.~\cite{campbell2001widths,perkins1991tables}. All core levels up to and including the Pt~4\textit{s} can be accessed with the SXPS photon energy. The error associated with the line width is $\pm0.5$~eV.}
    \label{fig:linewidths}
\end{figure}

As highlighted in previous work on W metal,~\cite{kalha2022lifetime} a vital aspect to consider when using core levels of different orbital nature, arising at different BEs, is the differences in lifetime broadening. Lifetime broadening arises from the creation of a core hole during photoemission. Core-level line widths in XPS contain contributions from Gaussian and Lorentzian components, wherein the former is generally attributed to non-lifetime effects such as instrumental factors, temperature, and vibrational broadening.~\cite{lee2023phase} Figure~\ref{fig:linewidths} compares the FWHM for all core levels measured using SXPS and HAXPES in this work to experimental and theoretical line widths obtained from Campbell~\textit{et al}.,~\cite{campbell2001widths} and theoretical line widths provided by Perkins~\textit{et al}.,~\cite{perkins1991tables} of which the summation of radiative and nonradiative line widths has been plotted. For SXPS, only core levels up to and including Pt~4\textit{s} can be accessed.\par
 
The Figure illustrates the large variation in the line widths of different core levels, with the largest difference observed between the Pt~4\textit{f} (0.60~eV) and Pt~3\textit{s} (13.7~eV) core levels, as measured by HAXPES. As seen for W, for orbitals with the same principal quantum number, $n$, the line width decreases as the orbital angular momentum, \textit{l}, increases, i.e. going from (\textit{l} = 0) \textit{s}- to (\textit{l = 3}) \textit{f}-orbitals. This can be attributed to reductions in the Coster-Kronig-Auger decay.~\cite{fuggle1980core} From the measured XPS presented in this work, the core levels agree better with the line widths from Ref.~\cite{perkins1991tables} than with those from Ref.~\cite{campbell2001widths}. Differences between the line widths from Ref.~\cite{campbell2001widths} and this work can be attributed to the calculation of the recommended `line width' calculated by Campbell~\textit{et al}. This involves averaging line widths measured from various experimental methods, including XPS, X-ray absorption spectroscopy (XAS), and X-ray emission spectroscopy (XES). Line widths derived in this work from SXPS and HAXPES data report similar values except for 5\textit{s}, where the line widths from SXPS and HAXPES are 14~eV and 6.5~eV, respectively. This is likely due to the differences in the ratios of $\frac{\sigma_i(Pt~4\textit{f})}{\sigma_i(Pt~5\textit{s})}$ at the two photon energies affecting the determination of line width for 5\textit{s}.\par

As discussed for W, another period 6 transition metal,~\cite{kalha2022lifetime} like Pt, the 3\textit{d} core level appears as a tractable alternative core level in HAXPES measurements, which can be used for studies that complement results obtained from the shallow Pt~4\textit{d} and Pt~4\textit{f} levels. This is due to the lower natural line width of the 3\textit{d} orbital compared to that of the 4\textit{d} orbital, making the determination of closely spaced chemical states easier. It could be argued that the shallow Pt~4\textit{f} with the lowest line width is the most suitable, but due to the close proximity of Pt~4\textit{f} to the $n$ = 5 core levels, determining different chemical states is more complicated than for Pt~3\textit{d}. Pt~3\textit{d} has a much larger $\Delta_\textrm{SOS}$ 80.2~eV compared to 4\textit{f} (3.4~eV), allowing for easy disentanglement of new peak environments for Pt and its related compounds. Additionally, in the photon energy regime of HAXPES (here 5.9~keV), the photoionisation cross-sections of Pt~3\textit{d} (0.04~Mb) are much higher than Pt~4\textit{f} (0.01~Mb), making acquisition of the more intense Pt~3\textit{d} far easier, especially in lab-based HAXPES systems where typical photon flux ranges are much lower than at synchrotron facilities.~\cite{spencer2021inelastic} \par

\subsection{Comparison of Core Level Satellites}\label{sec:CLcompare}

\begin{figure}[htbp]
\centering
    \includegraphics[keepaspectratio, width =0.8\linewidth]{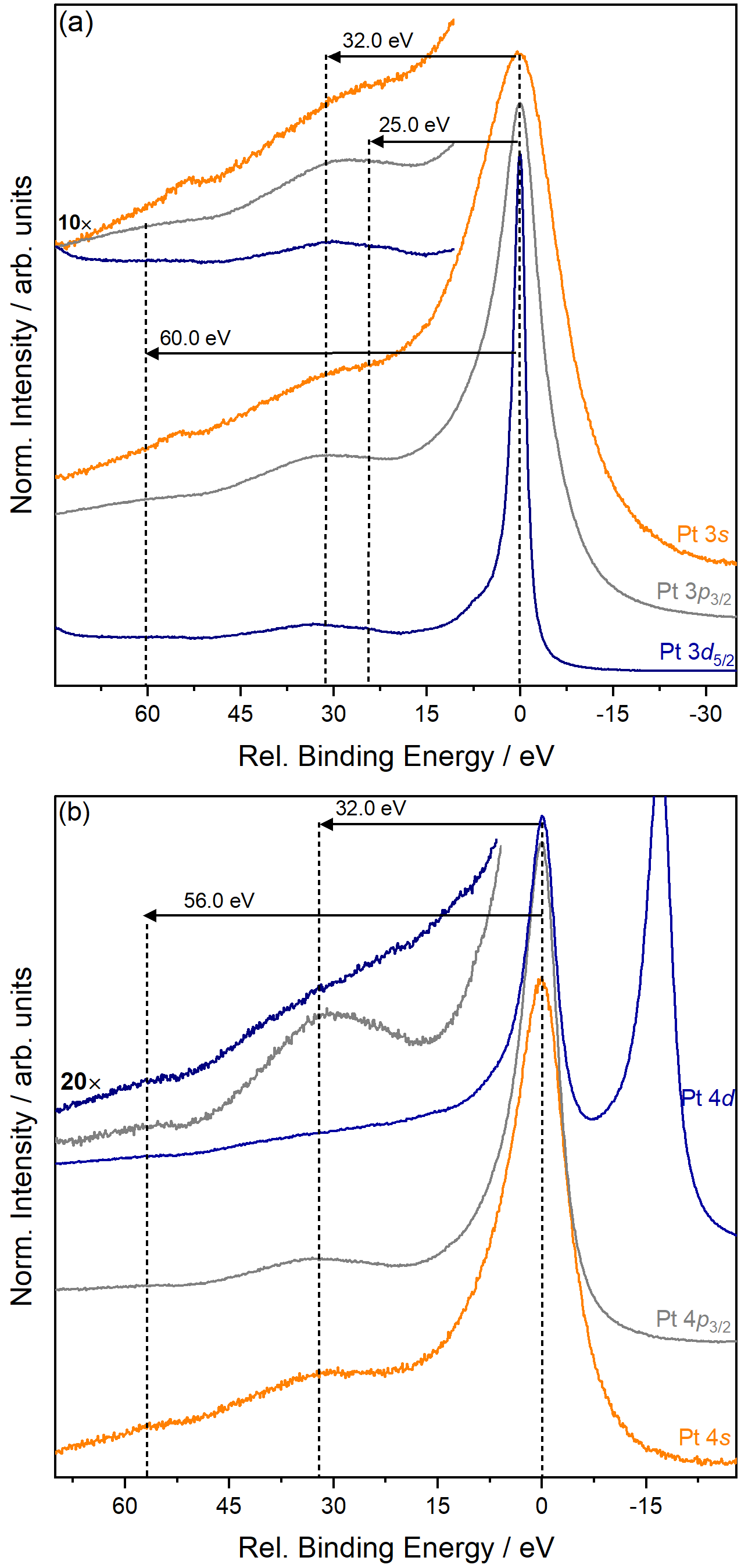}
    \caption{Comparison of the (a) Pt~3\textit{s}, Pt~3\textit{p}\textsubscript{3/2}, and Pt~3\textit{d}\textsubscript{5/2}, and (b) Pt~4\textit{s}, Pt~4\textit{p}\textsubscript{3/2}, and Pt~4\textit{d} core level HAXPES spectra on relative binding energy scales. Spectra are offset vertically, normalised to their maximum intensity, and aligned relative to the main photoionisation peak at 0~eV. Insets show expanded views of the higher binding energy tails.}
    \label{fig:ptcomp}
\end{figure}

When comparing the core-level spectra collected using SXPS and HAXPES, the satellite features observed correspond to similar features noted in the RHEEL spectrum. Figure~\ref{fig:ptcomp}(a) shows a comparison between the Pt~3\textit{s}, 3\textit{p}\textsubscript{3/2}, and 3\textit{d}\textsubscript{3/2} core spectra. The presence of the interband transition / bulk plasmon is noted for all three core levels, at $\approx32$~eV from the main photoionisation peak. The surface plasmon noted at 25.0~eV in the 3\textit{d} spectrum is also observed for 3\textit{p}\textsubscript{3/2}, but is difficult to observe in 3\textit{s} due to the lifetime broadening of the 3\textit{s} state affecting observation of structures at the higher energy tail. Additionally, the core levels also display the bulk plasmonic overtone seen at \textit{ca.}\ 60~eV in Pt~3\textit{p}\textsubscript{3/2} and Pt~3\textit{d}\textsubscript{5/2} (see the expanded inset in Figure~\ref{fig:ptcomp}(a)). Again, this is not seen for Pt~3\textit{s} due to the loss originating from 5\textit{p} states at $\approx56$~eV.

Similarly, Figure~\ref{fig:ptcomp}(b) shows the Pt~4\textit{s}, 4\textit{p}\textsubscript{3/2}, and 4\textit{d} core levels displaying the interband transition / bulk plasmon at 32~eV. This feature is difficult to see even in the inset for 4\textit{d} due to satellite overlap arising from both quasiparticle peaks, as described previously. The inset in Figure~\ref{fig:ptcomp}(b) shows that Pt~4\textit{p}\textsubscript{3/2} has a more intense contribution from the surface plasmon. The surface plasmon feature is most evidently seen for the \textit{p} orbitals, see Figure~\href{SI.pdf#fig:fig:ptspcomp}{S9} and~\href{SI.pdf#fig:Survey_hx}{S7} in the Supplementary Information for the comparison plot between Pt~n\textit{s} and \textit{p} core levels (n = 3, 4). Another feature observed for the n = 4 core levels is the loss associated with 5\textit{p} states around 56.0~eV. Across different core levels, the presence of satellite features in HAXPES and SXPS is consistent with the understanding of electronic losses from RHEEL measurements. The observation of some loss features appears to depend on the orbital angular momentum, which aids in understanding satellite structures in core-level spectra, and care must be taken to account for their presence during spectral fitting of XPS data.

\subsection{Valence Electronic Structure}\label{sec:Electronic}

Early work on the valence electronic structure of Pt formed part of broader investigations of various \textit{d}-state metals.~\cite{fadley1970electronic,clarke1974valence,kowalczyk1972high,baer1970band} Of these, Fadley and Shirley~\cite{fadley1970electronic} showed that the VB of 5\textit{d} metals are heavily influenced by SOC due to the strongly bound nature of the \textit{d}-states. They attributed SOC towards the shape of the Pt VB with two distinct features, as predicted theoretically. Similarly, Smith~\textit{et al}.\ used Al~K$\alpha$ X-ray excitation energy to examine the VB and compare it to DFT,~\cite{smith1974photoemission}, noting that the DFT-derived density of states (DOS) with SOC neglected resulted in a poor description of the experimental VB. Inclusion of SOC significantly influences \textit{d}-state contribution, improving VB comparisons. Both works concluded that SOC improves the description of the electronic state in transition metals, and therefore, these aspects are explored here with both DFT and G\textsubscript{0}W\textsubscript{0} approaches.\par 

\begin{figure}[htbp]
\centering
\includegraphics[keepaspectratio, width = 0.75\linewidth]{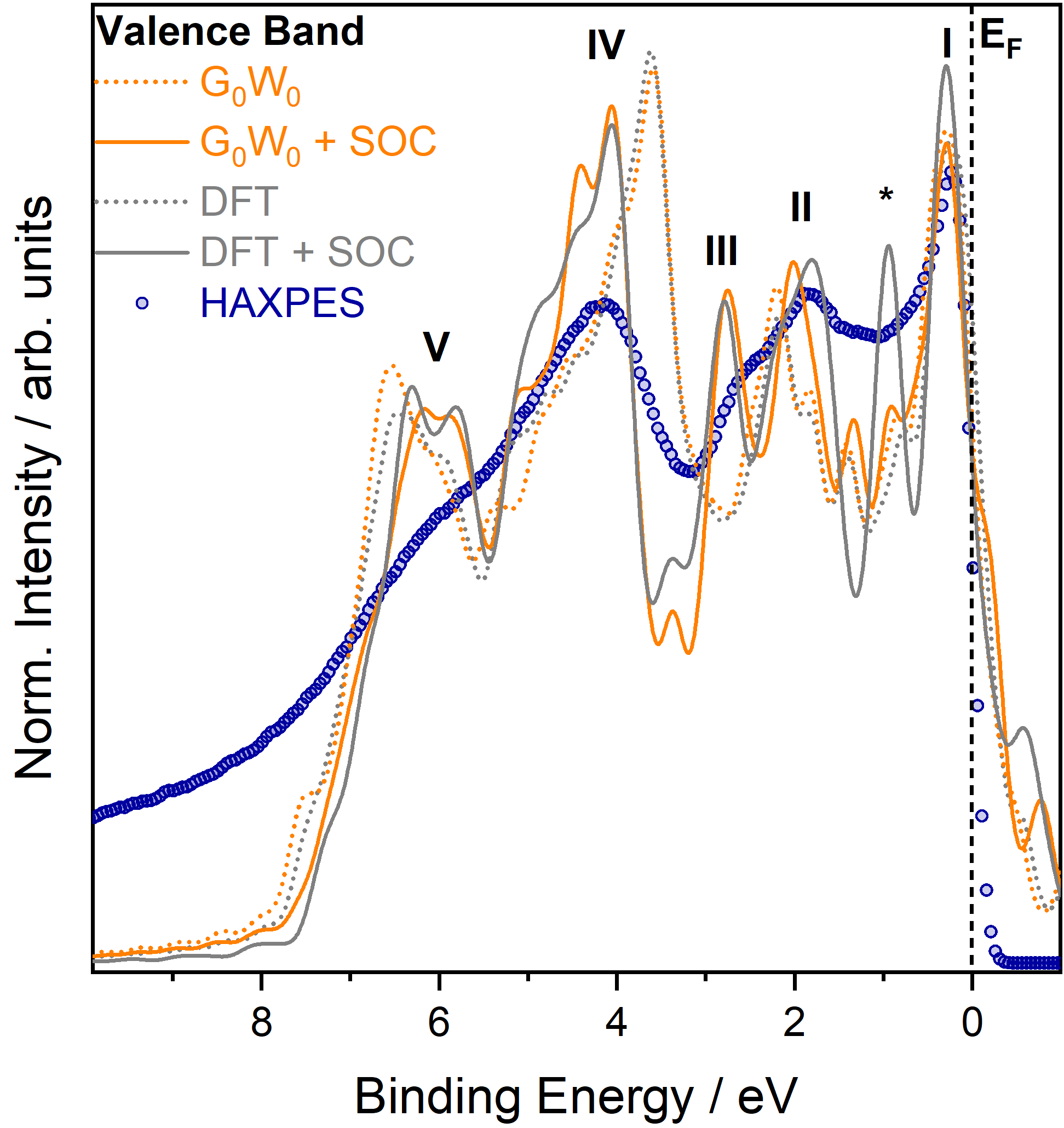}
    \caption{Comparison of the sum of individual PDOS after the Pb correction weighting approach from DFT and G\textsubscript{0}W\textsubscript{0} calculations with (solid lines) and without (dotted lines) the inclusion of SOC, along with HAXPES valence band
spectra. Roman numerals denote key features observed in the experimental spectra. The asterisk indicates a single feature not experimentally observable.}
    \label{fig:TDOS}
\end{figure}

Figure~\ref{fig:TDOS} shows the sum of the PDOS of Pt from four theoretical calculations, DFT and G\textsubscript{0}W\textsubscript{0}, each performed with and without SOC inclusion. The theoretical results are aligned to the E\textsubscript{F} position from HAXPES (<0.1~eV), with a Gaussian smearing equivalent to the total experimental resolution from HAXPES (0.258~eV) applied. The photoionisation cross-section weighting approach in Figure~\ref{fig:TDOS} is based on the Pb correction method (Approach 2 as outlined in the Methods Section). The Roman numerals correspond to key features seen in the experimental VB, which are discussed in detail in the following. 

All four PDOS summations show similar envelopes with two broad features present, one spanning 0-3.5~eV and the other 3.5-8~eV, consistent with the literature discussed above. The energy positions of most features I-V are similar, indicating all calculations can provide a reasonable, basic description of the valence electronic structure. The unweighted PDOS (see Figure~\href{SI.pdf#fig:DOS}{S2} of the Supplementary Information) also supports this, with similar BE positions and relative orbital contributions of features I-V. With the cross-section weighting, however, the relative intensity of features at 1~eV (*), 2~eV (II), and 2.7~eV (III), differ greatly between G\textsubscript{0}W\textsubscript{0} and DFT, both with and without SOC considerations.

From an initial comparison of the varying theoretical approaches with the experimental VB spectra, the two features for which the description markedly improves upon inclusion of SOC, regardless of theory level, are features III and IV. Firstly, feature III is essentially absent without the inclusion of SOC, whilst this appears as a clearly discernible peak in the experimental spectra. Secondly, the relative energy position of feature IV increases upon inclusion of SOC, providing a much better description of the observed experimental feature. The only discrepancy between the DFT + SOC and G\textsubscript{0}W\textsubscript{0} + SOC PDOS is the appearance of a peak at 1~eV (marked with an asterisk in Figure~\ref{fig:TDOS}) in the DFT-level calculation, which is not observed in experiment. Based on these observations, G\textsubscript{0}W\textsubscript{0} + SOC provides the best description of the experimental VB spectra in both SXPS and HAXPES. Therefore, it is used for the further exploration of the most optimal \textit{p}-state cross section approach. 

To obtain a complete understanding of the experimental VB, the 6\textit{p} contributions are explored with different photoionisation cross-section correction approaches, applied to the PDOS, and compared to the SXPS and HAXPES VB. These are shown in  Figures~\ref{fig:HX_final}, and ~\href{SI.pdf#fig:HX_SI}{S10} to ~\href{SI.pdf#fig:SX_SOC_final}{S13} of the Supplementary Information. All Subfigures (a-d) in Figure~\ref{fig:HX_final} show the PDOS with cross-section weighting using approaches (1)-(4) described in the Methods Section, with the relative cross-section values for all approaches at the HAXPES photon energy tabulated in Table~\ref{Xsections}. The experimental HAXPES data in Figure~\ref{fig:HX_final} show five distinct features, as noted for the summation of PDOS above. Features are located at 0.3~eV (I), 1.8~eV (II), 2.5~eV (III), 4.2~eV (IV), and 5.9~eV (V), agreeing with previously published data for the VB of a single crystal of Pt collected using $h\nu = $ Al K$\alpha$ (0.8~eV, 1.8~eV, 3.3~eV, 4.3~eV, and 6.1~eV).~\cite{ueda2022polarization,wu2020electronic,kowalczyk1972high}\par 

\begin{figure*}
\centering
    \includegraphics[keepaspectratio, width = 0.6\linewidth]{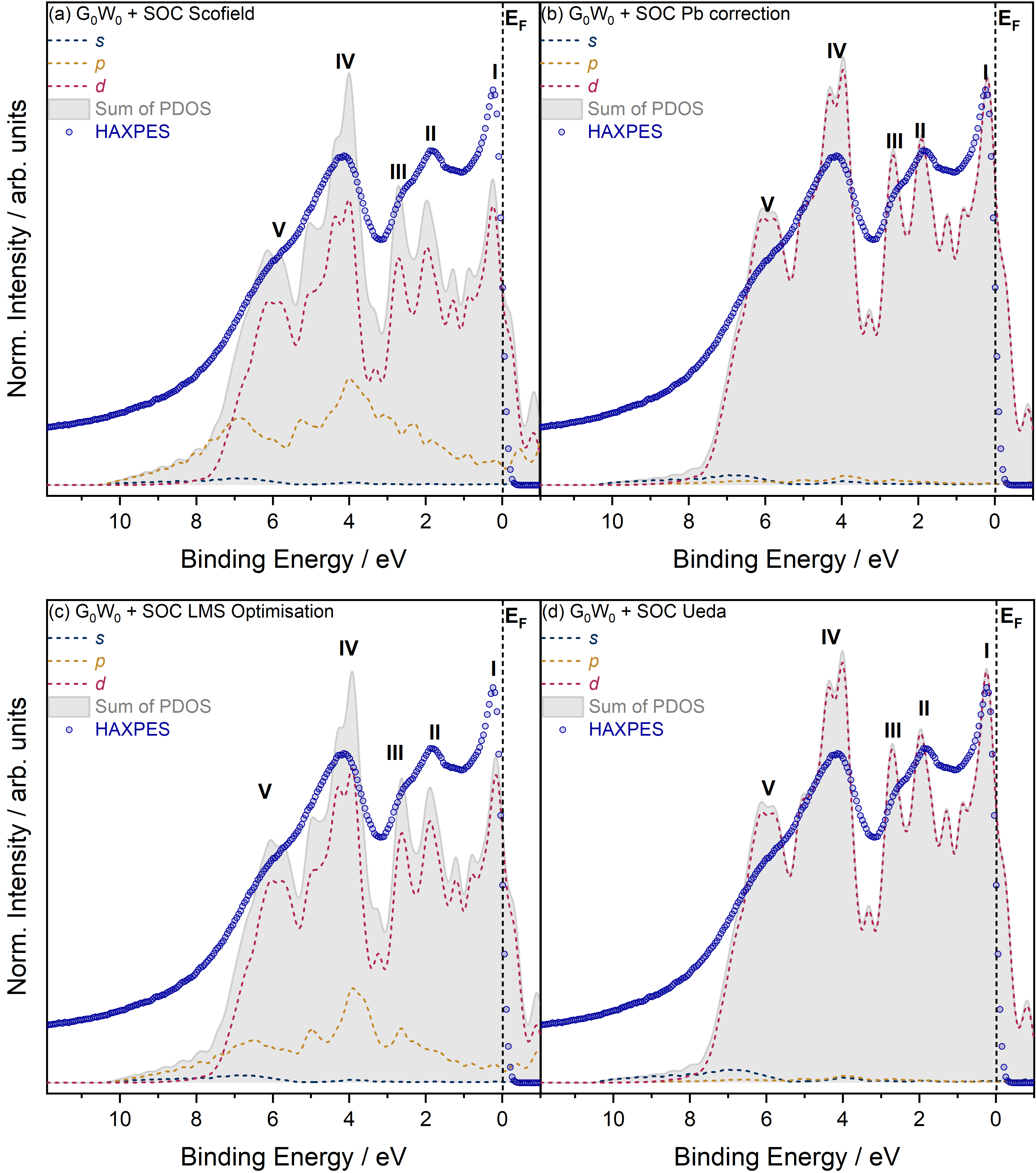}
    \caption{Comparison of the PDOS spectra calculated using G\textsubscript{0}W\textsubscript{0} and SOC with HAXPES valence band spectra, including the (a) Scofield (approach 1), (b) Pb correction (approach 2), (c) LMS optimisation (approach 3), and (d) Ueda (approach 4) \textit{p}-state cross-section weighting approaches. The PDOS contributions have been broadened to match the experimental broadening. HAXPES spectra are normalised to their respective areas after removing a Shirley-type background.}
    \label{fig:HX_final}
\end{figure*}

\begin{table}[!htbp]
    \caption{\label{Xsections}A summary of the one-electron photoionisation cross-sections $\sigma_i$ relative to the 5\textit{d} $\sigma_i$ taken from various cross-section weighting approaches, including (1) Scofield, (2) Pb correction, (3) Least-mean square (LMS), and (4) Ueda from the HAXPES spectrum. The relative cross-sections for approach (3) LMS have been taken from the optimisation of the G\textsubscript{0}W\textsubscript{0} calculation. The values obtained from approach (4) Ueda~\textit{et al.} were determined at 6.0~keV using horizontal (H) polarised light.~\cite{ueda2022polarization}  The values listed in the remaining rows were extracted from the Scofield photoionisation cross-section data, and the Galore software package was used to interpolate the data and estimate the photoionisation cross-section at the hard X-ray energy used in this work (5.9266~keV).~\cite{j2018galore}}
    \begin{ruledtabular}
    \begin{tabular}{cccc}
    Approach & \textbf{5\textit{d}} & \textbf{6\textit{s}} & \textbf{6\textit{p}} \\
    \hline
    (1) Scofield & 1.00 & 0.2934 & 3.303 \\
    (2) Pb correction & 1.00 & 0.2934 & 0.1735  \\
    (3) LMS & 1.00 & 0.2934 & 2.3690  \\
    (4) Ueda & 1.00 & 0.3845 & 0.1289  \\
    \end{tabular}
    \end{ruledtabular}
\end{table}

Since BE positions of features are similar in Figure~\ref{fig:HX_final}, the relative intensities of the features are considered, as this is directly influenced by the level of \textit{p}-orbital contribution. The majority of VB contributions arise from 5\textit{d} orbitals, with the 6\textit{sp} bands providing a much smaller contribution, as is expected from the electronic structure of Pt. All features are a result of a mixing of the 5\textit{d} and 6\textit{sp} states, where the \textit{d} orbitals dominate. While the 6\textit{p} states are distributed across the entire width of VB, the 6\textit{s} contributions are more localised at the bottom of the VB (\textit{ca.} 4 - 7~eV), influencing features IV and V. Concerning these features, it is noted that in all approaches, the intensity of features IV and V are overestimated. This is attributed to final-state effects, in particular non-constant lifetime broadening, which are not considered in the DOS calculations but are observed in XPS and have been noted previously.~\cite{kalha2022lifetime} However, since the relative intensity between features IV and V in the DOS remains constant and corresponds well to the same features in the HAXPES VB, the contributions of the \textit{d} and \textit{s} states for those features remain valid.\par

Looking then at the effect of the 6\textit{p} contribution, the PDOS predicts that \textit{p} character is most evident in feature IV and lowest at feature I, demonstrating the highest and lowest \textit{p} state contributions, respectively. The Scofield (1, Figure~\ref{fig:HX_final}(a)) and LMS optimisation (3, Figure~\ref{fig:HX_final}(c)) approaches show significant \textit{p}-state orbital contribution, while the Pb correction (2, Figure~\ref{fig:HX_final}(b)) and Ueda (4, Figure~\ref{fig:HX_final}(d)) approaches indicate minimal \textit{p} orbital presence. The relative intensity between features I and IV, representing the lowest and highest 6\textit{p} densities, can be used to determine the quality of the VB description by the DOS.  Where the 6\textit{p} contribution is significant, i.e. Approaches (1) and (3), the relative intensity of I and IV in the DOS is much larger than in the experimental VB, resulting in a much poorer description. This indicates that \textit{p}-state contribution after cross sections are taken into account is likely minimal, similar to 6\textit{s} (see Table~\ref{Xsections}). Approaches (2) and (4) with similar relative 6\textit{s} and 6\textit{p} cross sections (with relative contributions between 0.1-0.4) show a much better intensity of I compared to experiment. Hence, the order of poorest to best descriptions of the HAXPES VB are Scofield (1), LMS (3), Ueda (4), and Pb correction (2). Approach (1) uses the 5\textit{p} cross-section weighting, resulting in feature I having the lowest relative intensity compared to IV. Additionally, relative intensities for features II and III also deviate from experiment, showing lower II than III intensity in the DOS. While Approach (3) shows the same, the relative intensity of I and IV is better. Comparing the Pb correction (2) and Ueda (4) approaches both show the correct relative intensity trends for all features, but Ueda (4) slightly overestimates the relative intensity of feature I, making the Pb correction (2) the best approach for describing the VB.\par 

\begin{figure}[htbp]
\centering
\includegraphics[keepaspectratio, width = 0.75\linewidth]{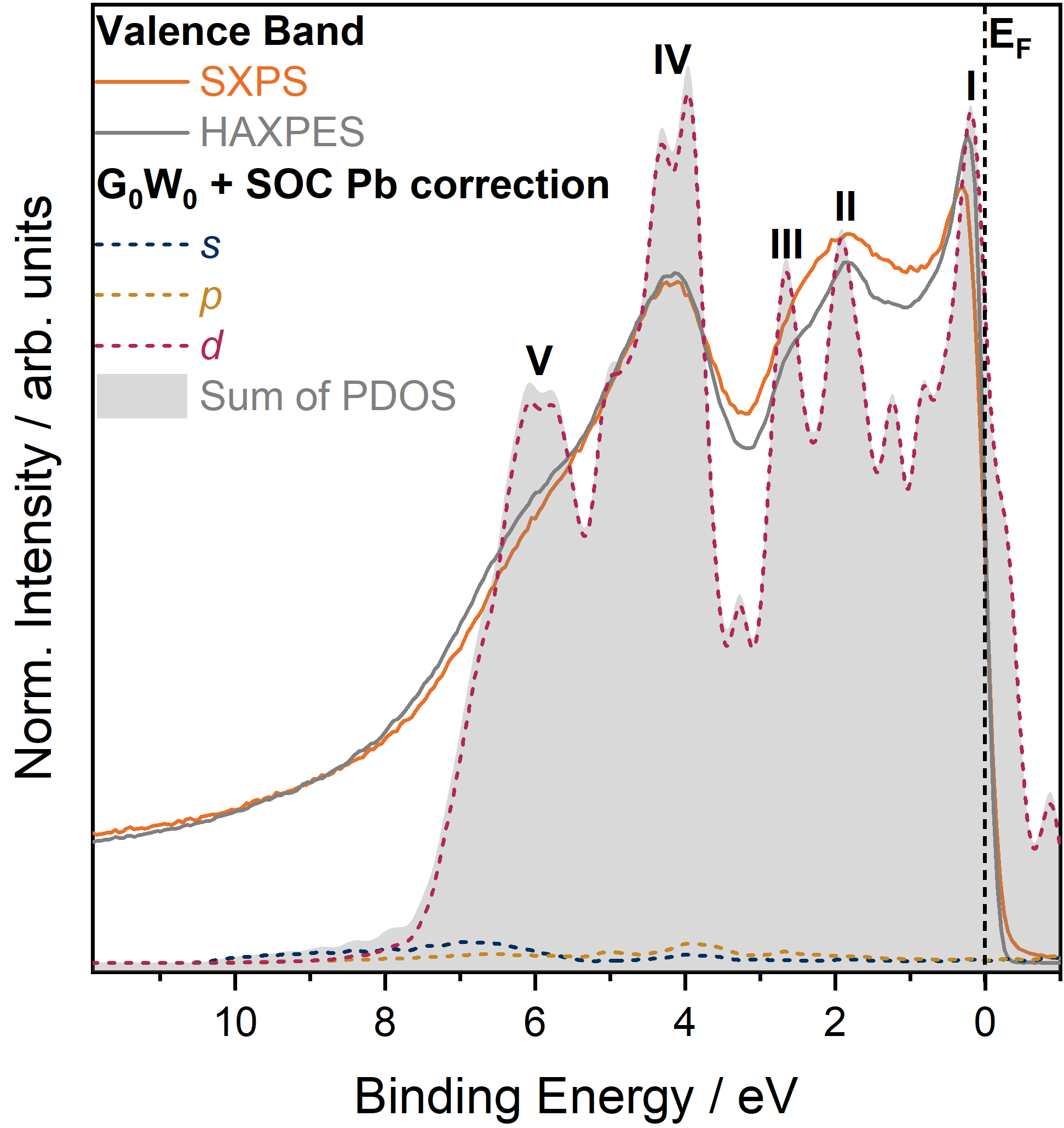}
    \caption{Comparison of the PDOS spectra calculated using G\textsubscript{0}W\textsubscript{0} and SOC with synchrotron-based SXPS ($h\nu$ = 1.7~keV) and HAXPES ($h\nu$ = 5.9~keV). Experimental spectra are normalised to the total area after removal of a Shirley-type background. In the PDOS, dotted lines depict the contributions of individual PDOS after the Pb correction cross-section weighting approach was applied, assuming the HAXPES photon energy.}
    \label{fig:VB}
\end{figure}

These observations also apply to VB from SXPS seen in Figure~\ref{fig:VB}, which displays the VB spectra from SXPS and HAXPES with the DOS calculated from G\textsubscript{0}W\textsubscript{0} + SOC, weighted with approach (2) at the HAXPES photon energy. Given the dominance of $d$ states in the VB, and the comparability of the energy resolution achieved in the experiment (both $<$300~meV), it is not surprising that the experimental spectra collected with both photon energies are highly comparable. Subtle differences, particularly around features II and III, can be attributed to the differences in photoionisation cross-sections for the $p$ states, further confirming the accuracy of the theoretical description and cross-section correction approach chosen.\par

Overall, both DFT and G\textsubscript{0}W\textsubscript{0} approaches are able to produce a good description of the valence DOS of Pt as observed in experiment. In both, the consideration of spin-orbit coupling improves the envelope of the DOS in comparison to experimental VB spectra significantly, in particular for features III and IV, with G\textsubscript{0}W\textsubscript{0} + SOC providing a very good description of the experimental spectra, both in terms of relative energy positions and intensities.

\section{Conclusions}

This work presents a comprehensive spectroscopic and theoretical investigation of metallic platinum, combining RHEELS, SXPS, HAXPES, and \textit{ab initio} electronic-structure calculations to establish a unified description of its photoemission response. The measurements reveal pronounced metallic line-shape asymmetries and extensive satellite structure throughout the Pt photoelectron spectrum, demonstrating the importance of many-body final-state effects across both shallow and deep core levels.

By correlating photoelectron satellites with independently measured RHEELS loss features, we directly connect satellite formation in Pt to specific energy-loss processes, including interband transitions, surface and bulk plasmons, plasmonic overtones, and semi-core excitation channels. This analysis enables the assignment of several previously unresolved spectral features and provides, to our knowledge, the first internally consistent interpretation of satellite structure across the accessible Pt core levels. The study additionally establishes experimental values for quantities that have received limited attention in the literature, including the Pt 5\textit{p} spin–orbit splitting and systematic trends in core-hole lifetime broadening.

Comparison of experimental VB spectra with DFT- and G\textsubscript{0}W\textsubscript{0}-derived PDOS demonstrates that an accurate description of Pt requires both SOC and appropriate treatment of the 6\textit{p}-state contribution. Among the approaches considered, G\textsubscript{0}W\textsubscript{0} calculations including SOC provide the closest agreement with experiment, reproducing the observed fine structure of the valence electronic states.

Taken together, these results provide a critically assessed spectroscopic benchmark for metallic platinum and clarify the relationship between its electronic structure and many-body photoemission processes. The reference data and assignments reported here should facilitate more rigorous interpretation of XPS and HAXPES measurements of Pt-containing catalysts, electronic materials, and related 5\textit{d} transition-metal systems.

\begin{acknowledgments}
AAR acknowledges the support from the Department of Chemistry, UCL. We acknowledge Diamond Light Source for time on Beamline I09 under Proposals SI34325-1 and SI36180-4. We acknowledge DESY (Hamburg, Germany), a member of the Helmholtz Association HGF, for the provision of experimental facilities. Parts of this research were carried out at PETRA III, beamline P22. Beamtime was allocated for proposal I-20230848. REELS was carried out at HarwellXPS, the UK National XPS Facility (EP/Y023587/1). J.J.G.M. acknowledges the AI4S fellowship within the “Generación D” initiative by Red.es, Ministerio para la Transformación Digital y de la Función Pública, for talent attraction (C005/24-ED CV1), funded by NextGenerationEU through the Recovery, Transformation and Resilience Plans (PRTR). We also thank the EUROPRACTICE network for providing us with access to modelling software at a reasonable cost.
\end{acknowledgments}

\section*{Disclosures}
The authors declare no conflicts of interest.

\section*{Data Availability}
Data for this article, including all processed data of the main paper figures, are available at Zenodo in Origin format at [\href{https://doi.org/10.5281/zenodo.17295609}{https://doi.org/10.5281/zenodo.17295609}].

\section*{Supplementary Information}
Further details on the experimental electron energy-loss spectroscopy can be found in the Supplementary Information. Additionally, unweighted DOS from the DFT and G\textsubscript{0}W\textsubscript{0} calculations, with and without SOC, are provided. Lastly, survey spectra, comparisons of satellite features at other core levels, and all plots of the different \textit{p}-orbital approaches used to determine the optimum photoionisation cross-section weighting for all theoretical calculation methods can also be found in the Supplementary Information.

\bibliographystyle{apsrev4-1}
\bibliography{references.bib}

@article{ueda2022polarization,
  title={Polarization-dependent Bulk-sensitive Valence Band Photoemission Spectroscopy and Density Functional Theory Calculations: Part III. 5 d Transition Metals},
  author={Ueda, Shigenori and Hamada, Ikutaro},
  journal={Journal of the Physical Society of Japan},
  volume={91},
  number={2},
  pages={024801},
  year={2022},
  publisher={The Physical Society of Japan}
}

@article{resende2011effect,
  title={The effect of coating TiO 2 on the CO oxidation of the Pt/$\gamma$-alumina catalysts},
  author={Resende, Neuman S and Perez, Carlos A and Eon, Jean G and Schmal, Martin},
  journal={Catalysis Letters},
  volume={141},
  pages={1685--1692},
  year={2011},
  publisher={Springer}
}

@TechReport{Scofield1973,
author={Scofield, J. H.},
title={{Theoretical photoionization cross sections from 1 to 1500 keV.}},
institution ={{University of California, Lawrence Livermore Laboratory}},
year={1973},
address={United States},
doi={10.2172/4545040},
url={https://doi.org/10.2172/4545040},
}

@misc{Dig_Sco_2020,
    title = {{Digitisation of Scofield Photoionisation Cross Section Tabulated Data}},
    year = {2020},
    author = {Kalha, Curran and Fernando, Nathalie K and Regoutz, Anna},
    url = {https://figshare.com/articles/dataset/Digitisation_of_Scofield_Photoionisation_Cross_Section_Tabulated_Data/12967079},
    doi = {10.6084/m9.figshare.12967079.v1}
}

@misc{Kalha2020,
author = "Curran Kalha and Nathalie Fernando and Anna Regoutz",
title = "{Digitisation of Scofield Photoionisation Cross Section Tabulated Data}",
year = "2020",
month = "9",
url = "https://figshare.com/articles/dataset/Digitisation_of_Scofield_Photoionisation_Cross_Section_Tabulated_Data/12967079",
doi = "10.6084/m9.figshare.12967079.v1"
}

@misc{Dig_Pt,
    title = {{HAXPES spectra of Pt Sputtered}},
    year = {2020},
    author = {Satoshi Yasuno},
    url = {https://mdr.nims.go.jp/concern/datasets/hx11xj53m},
    doi = {10.48505/nims.3275}
}

@article{wise1950platinum,
  title={The platinum metals: a review of their properties and uses},
  author={Wise, Edmund M},
  journal={Journal of The Electrochemical Society},
  volume={97},
  number={3},
  pages={57C},
  year={1950},
  publisher={IOP Publishing}
}

@article{yu2012review,
  title={Review of Pt-based bimetallic catalysis: from model surfaces to supported catalysts},
  author={Yu, Weiting and Porosoff, Marc D and Chen, Jingguang G},
  journal={Chemical Reviews},
  volume={112},
  number={11},
  pages={5780--5817},
  year={2012},
  publisher={ACS Publications}
}

@article{an2014design,
  title={Design of a highly active Pt/Al 2 O 3 catalyst for low-temperature CO oxidation},
  author={An, Nihong and Yuan, Xiaoling and Pan, Bo and Li, Qinglin and Li, Suying and Zhang, Wenxiang},
  journal={RSC Advances},
  volume={4},
  number={72},
  pages={38250--38257},
  year={2014},
  publisher={Royal Society of Chemistry}
}

@article{he2024single,
  title={Single-Atom Alloys Materials for CO2 and CH4 Catalytic Conversion},
  author={He, Chengxuan and Gong, Yalin and Li, Songting and Wu, Jiaxin and Lu, Zhaojun and Li, Qixin and Wang, Lingzhi and Wu, Shiqun and Zhang, Jinlong},
  journal={Advanced Materials},
  volume={36},
  number={16},
  pages={2311628},
  year={2024},
  publisher={Wiley Online Library}
}

@article{akhtar2022improvement,
  title={The improvement in surface properties of metallic implant via magnetron sputtering: recent progress and remaining challenges},
  author={Akhtar, Memoona and Uzair, Syed Ahmed and Rizwan, Muhammad and Ur Rehman, Muhammad Atiq},
  journal={Frontiers in Materials},
  volume={8},
  pages={747169},
  year={2022},
  publisher={Frontiers Media SA}
}

@article{johnsson2022review,
  title={A review of platinum diffusion in silicon and its application for lifetime engineering in power devices},
  author={Johnsson, Anna and Schmidt, Gerhard and Hauf, Moritz and Pichler, Peter},
  journal={Physica Status Solidi A},
  volume={219},
  number={2},
  pages={2100462},
  year={2022},
  publisher={Wiley Online Library}
}

@article{wach2020comparative,
  title={Comparative study of the around-Fermi electronic structure of 5d metals and metal-oxides by means of high-resolution X-ray emission and absorption spectroscopies},
  author={Wach, Anna and S{\'a}, Jacinto and Szlachetko, Jakub},
  journal={Synchrotron Radiation},
  volume={27},
  number={3},
  pages={689--694},
  year={2020},
  publisher={International Union of Crystallography}
}

@article{isaacs2021advanced,
  title={Advanced XPS characterization: XPS-based multi-technique analyses for comprehensive understanding of functional materials},
  author={Isaacs, Mark A and Davies-Jones, Josh and Davies, Philip R and Guan, Shaoliang and Lee, Roxy and Morgan, David J and Palgrave, Robert},
  journal={Materials Chemistry Frontiers},
  volume={5},
  number={22},
  pages={7931--7963},
  year={2021},
  publisher={Royal Society of Chemistry}
}

@article{mason2023platinum,
  title={Platinum-based chemotherapy for early triple-negative breast cancer},
  author={Mason, Sofia RE and Willson, Melina L and Egger, Sam J and Beith, Jane and Dear, Rachel F and Goodwin, Annabel},
  journal={Cochrane Database of Systematic Reviews},
  number={9},
  year={2023},
  publisher={John Wiley \& Sons, Ltd}
}

@article{rong2021pt,
  title={Pt single atom-induced activation energy and adsorption enhancement for an ultrasensitive ppb-level methanol gas sensor},
  author={Rong, Qian and Xiao, Bin and Zeng, Jiyang and Yu, Ruohan and Zi, Baoye and Zhang, Genlin and Zhu, Zhongqi and Zhang, Jin and Wu, Jinsong and Liu, Qingju},
  journal={ACS Sensors},
  volume={7},
  number={1},
  pages={199--206},
  year={2021},
  publisher={ACS Publications}
}

@article{jiao2023enhancing,
  title={Enhancing the temperature coefficient of resistance of Pt thin film resistance-temperature-detector by short-time annealing},
  author={Jiao, Ruina and Wang, Kunlun and Xin, Yanqing and Sun, Hui and Gong, Jianhong and Yu, Lan and Wang, Yong},
  journal={Ceramics International},
  volume={49},
  number={8},
  pages={12596--12603},
  year={2023},
  publisher={Elsevier}
}

@article{jin2024development,
  title={Development of platinum complexes for tumor chemoimmunotherapy},
  author={Jin, Suxing and Guo, Yan and Wang, Xiaoyong},
  journal={Chemistry--A European Journal},
  volume={30},
  number={10},
  pages={e202302948},
  year={2024},
  publisher={Wiley Online Library}
}

@article{vovk2017xps,
  title={XPS study of stability and reactivity of oxidized Pt nanoparticles supported on TiO2},
  author={Vovk, Evgeny I and Kalinkin, Alexander V and Smirnov, Mikhail Yu and Klembovskii, Igor O and Bukhtiyarov, Valerii I},
  journal={The Journal of Physical Chemistry C},
  volume={121},
  number={32},
  pages={17297--17304},
  year={2017},
  publisher={ACS Publications}
}

@article{yang2023revealing,
  title={Revealing the Surface Species Evolution on Low-loading Platinum in an Electrochemical Redox Reaction by Operando Ambient-Pressure X-ray Photoelectron Spectroscopy},
  author={Yang, Chueh-Cheng and Tsai, Meng-Hsuan and Yang, Zong-Ren and Tseng, Yuan-Chieh and Wang, Chia-Hsin},
  journal={ChemCatChem},
  volume={15},
  number={12},
  pages={e202300359},
  year={2023},
  publisher={Wiley Online Library}
}

@article{siburian2021loading,
  title={The loading effect of Pt clusters on Pt/graphene nano sheets catalysts},
  author={Siburian, Rikson and Ali, Ab Malik Marwan and Sebayang, Kerista and Supeno, Minto and Tarigan, Kerista and Simanjuntak, Crystina and Aritonang, Sri Pratiwi and Hutagalung, Fajar},
  journal={Scientific Reports},
  volume={11},
  number={1},
  pages={2532},
  year={2021},
  publisher={Nature Publishing Group UK London}
}

@article{clarke1974valence,
  title={The valence band of platinum: Esca studies based on clean (100) and (111) surfaces and their chemisorption of carbon monoxide and other unsaturated molecules},
  author={Clarke, Terence A and Gay, Ian D and Mason, Ronald},
  journal={Chemical Physics Letters},
  volume={27},
  number={2},
  pages={172--174},
  year={1974},
  publisher={Elsevier}
}

@article{kowalczyk1972high,
  title={HIGH-RESOLUTION XPS SPECTRA OF Ir, Pt, AND Au VALENCE BANDS},
  author={Kowalczyk, S and Ley, L and Pollak, R and Shirley, DA},
  year={1972},
  journal={Physics Letters A},
}

@article{baer1970band,
  title={Band structure of transition metals studied by ESCA},
  author={Baer, Y and Heden, PF and Hedman, J and Klasson, M and Nordling, C and Siegbahn, K},
  journal={Physica Scripta},
  volume={1},
  number={1},
  pages={55},
  year={1970},
  publisher={IOP Publishing}
}

@article{j2018galore,
  title={Galore: Broadening and weighting for simulation of photoelectron spectroscopy},
  author={J Jackson, Adam and M Ganose, Alex and Regoutz, Anna and G Egdell, Russell and D Scanlon, O},
  journal={Journal of Open Source Software},
  volume={3},
  number={26},
  year={2018},
  publisher={The Open Journal}
}

@article{schlueter2019new,
  title={The new dedicated HAXPES beamline P22 at PETRAIII},
  author={Schlueter, C and Gloskovskii, A and Ederer, K and Schostak, I and Piec, S and Sarkar, I and Matveyev, Yu and L{\"o}mker, P and Sing, M and Claessen, R and others},
  journal={AIP Conference Proceedings},
  volume={2054},
  year={2019},
  organization={AIP Publishing}
}

@article{smith1974photoemission,
  title={Photoemission spectra and band structures of d-band metals. III. Model band calculations on Rh, Pd, Ag, Ir, Pt, and Au},
  author={Smith, Neville V},
  journal={Physical Review B},
  volume={9},
  number={4},
  pages={1365},
  year={1974},
  publisher={APS}
}

@article{wolstenholme2008summary,
  title={Summary of ISO/TC 201 Standard: XXX. ISO 18516: 2006—Surface chemical analysis—Auger electron spectroscopy and X-ray photoelectron spectroscopy—Determination of lateral resolution},
  author={Wolstenholme, J},
  journal={Surface and Interface Analysis: An International Journal devoted to the development and application of techniques for the analysis of surfaces, interfaces and thin films},
  volume={40},
  number={5},
  pages={966--968},
  year={2008},
  publisher={Wiley Online Library}
}

@article{Duncan_2018,
author = {Tien-Lin Lee and David A. Duncan},
title = {A Two-Color Beamline for Electron Spectroscopies at Diamond Light Source},
journal = {Synchrotron Radiation News},
volume = {31},
number = {4},
pages = {16-22},
year  = {2018},
publisher = {Taylor & Francis},
}

@article{tanuma1994calculations,
  title={Calculations of electron inelastic mean free paths. V. Data for 14 organic compounds over the 50--2000 eV range},
  author={Tanuma, Shigeo and Powell, Cedric J and Penn, David R},
  journal={Surface and Interface Analysis},
  volume={21},
  number={3},
  pages={165--176},
  year={1994},
  publisher={Wiley Online Library}
}

@article{mueller1971quadratic,
  title={Quadratic integration: theory and application to the electronic structure of platinum},
  author={Mueller, FM and Garland, JW and Cohen, MH and Bennemann, KH},
  journal={Annals of Physics},
  volume={67},
  number={1},
  pages={19--57},
  year={1971},
  publisher={Elsevier}
}

@article{wu2020electronic,
  title={On the electronic structure and hydrogen evolution reaction activity of platinum group metal-based high-entropy-alloy nanoparticles},
  author={Wu, Dongshuang and Kusada, Kohei and Yamamoto, Tomokazu and Toriyama, Takaaki and Matsumura, Syo and Gueye, Ibrahima and Seo, Okkyun and Kim, Jaemyung and Hiroi, Satoshi and Sakata, Osami and others},
  journal={Chemical Science},
  volume={11},
  number={47},
  pages={12731--12736},
  year={2020},
  publisher={Royal Society of Chemistry}
}

@article{froidevaux1968electronic,
  title={Electronic structure of platinum and palladium alloys},
  author={Froidevaux, C and Launois, H and Gautier, F},
  journal={Journal of Applied Physics},
  volume={39},
  number={2},
  pages={557--558},
  year={1968},
  publisher={American Institute of Physics}
}

@article{zhang2022platinum,
  title={Platinum-based drugs for cancer therapy and anti-tumor strategies},
  author={Zhang, Chunyu and Xu, Chao and Gao, Xueyun and Yao, Qingqiang},
  journal={Theranostics},
  volume={12},
  number={5},
  pages={2115},
  year={2022}
}

@article{ienco2023role,
  title={The Role of Inverted Ligand Field in the Electronic Structure and Reactivity of Octahedral Formal Platinum (IV) Complexes},
  author={Ienco, Andrea and Ruffo, Francesco and Manca, Gabriele},
  journal={Chemistry--A European Journal},
  volume={29},
  number={59},
  pages={e202301669},
  year={2023},
  publisher={Wiley Online Library}
}

@article{fadley1970electronic,
  title={Electronic densities of states from X-ray photoelectron spectroscopy},
  author={Fadley, CS and Shirley, DA},
  journal={Journal of Research of the National Bureau of Standards A},
  volume={74},
  number={4},
  pages={543},
  year={1970}
}

@article{anniyev2010complementarity,
  title={Complementarity between high-energy photoelectron and L-edge spectroscopy for probing the electronic structure of 5d transition metal catalysts},
  author={Anniyev, Toyli and Ogasawara, Hirohito and Ljungberg, Mathias P and Wikfeldt, Kjartan T and MacNaughton, Janay B and N{\"a}slund, Lars-{\AA}ke and Bergmann, Uwe and Koh, Shirlaine and Strasser, Peter and Pettersson, Lars GM and others},
  journal={Physical Chemistry Chemical Physics},
  volume={12},
  number={21},
  pages={5694--5700},
  year={2010},
  publisher={Royal Society of Chemistry}
}

@article{zheng2023haxpes,
  title={HAXPES reference spectra of Au and Pt with Cr K$\alpha$ excitation},
  author={Zheng, Dong and Young, Christopher N and Stickle, William F},
  journal={Surface Science Spectra},
  volume={30},
  number={2},
  year={2023},
  publisher={AIP Publishing}
}

@article{rumaiz2023interface,
  title={Interface formation and Schottky barrier height for Y, Nb, Au, and Pt on Ge as determined by hard x-ray photoelectron spectroscopy},
  author={Rumaiz, Abdul K and Weiland, Conan and Harding, Ian and Nooman, Neha S and Krings, Thomas and Hull, Ethan L and Giacomini, Gabriele and Chen, Wei and Cockayne, Eric and Siddons, D Peter and others},
  journal={AIP Advances},
  volume={13},
  number={1},
  year={2023},
  publisher={AIP Publishing}
}

@article{zborowski2022reference,
  title={Reference survey spectra of elemental solid measured with Cr K$\alpha$ photons as a tool for Quases analysis (3): Transition metals period 6 elements (Hf, Ta, W, Re, Ir, Pt, Au)},
  author={Zborowski, C and Conard, T and Vanleenhove, A and Hoflijk, I and Vaesen, I},
  journal={Surface Science Spectra},
  volume={29},
  number={2},
  year={2022},
  publisher={AIP Publishing}
}

@article{kalha2022lifetime,
  title={Lifetime effects and satellites in the photoelectron spectrum of tungsten metal},
  author={Kalha, Curran and Ratcliff, Laura E and Moreno, JJ Guti{\'e}rrez and Mohr, Stephan and Mantsinen, Mervi and Fernando, Nathalie K and Thakur, Pardeep K and Lee, T-L and Tseng, H-H and Nunney, Tim S and others},
  journal={Physical Review B},
  volume={105},
  number={4},
  pages={045129},
  year={2022},
  publisher={APS}
}

@article{powell1960origin,
  title={The origin of the characteristic electron energy losses in ten elements},
  author={Powell, CJ},
  journal={Proceedings of the Physical Society},
  volume={76},
  number={5},
  pages={593},
  year={1960},
  publisher={IOP Publishing}
}

@article{rudberg1930characteristic,
  title={Characteristic energy losses of electrons scattered from incandescent solids},
  author={Rudberg, Erik},
  journal={Proceedings of the Royal Society of London A},
  volume={127},
  number={804},
  pages={111--140},
  year={1930},
  publisher={The Royal Society London}
}

@article{pines1956collective,
  title={Collective energy losses in solids},
  author={Pines, David},
  journal={Reviews of modern physics},
  volume={28},
  number={3},
  pages={184},
  year={1956},
  publisher={APS}
}

@article{schroder1974investigation,
  title={Investigation of the characteristic energy losses of platinum (111)},
  author={Schr{\"o}der, W and Peters, E and H{\"o}lzl, J},
  journal={Applied Physics},
  volume={3},
  pages={135--140},
  year={1974},
  publisher={Springer}
}

@article{albert1956einfluss,
  title={{\"U}ber den Einflu{\ss} diskreter Energieverluste der Elektronen auf die Struktur der Bremsstrahl-Isochromaten},
  author={Albert, Ludwig},
  journal={Zeitschrift f{\"u}r Physik},
  volume={143},
  number={5},
  pages={513--532},
  year={1956},
  publisher={Springer}
}

@article{jaegle1969experimental,
  title={Experimental and theoretical study of the absorption of ultrasoft X rays in some heavy elements},
  author={Jaegl{\'e}, P and Farnoux, F Combet and Dhez, P and Cremonese, M and Onori, G},
  journal={Physical Review},
  volume={188},
  number={1},
  pages={30},
  year={1969},
  publisher={APS}
}

@article{haensel1969optical,
  title={Optical absorption measurements of tantalum, tungsten, rhenium and platinum in the extreme ultraviolet},
  author={Haensel, R and Radler, K and Sonntag, B and Kunz, C},
  journal={Solid State Communications},
  volume={7},
  number={20},
  pages={1495--1497},
  year={1969},
  publisher={Elsevier}
}

@article{melinon1997nanostructured,
  title={Nanostructured silicon films obtained by neutral cluster depositions},
  author={M{\'e}linon, P and K{\'e}gh{\'e}lian, P and Pr{\'e}vel, B and Perez, A and Guiraud, G and LeBrusq, J and Lerm{\'e}, J and Pellarin, M and Broyer, Mp},
  journal={The Journal of chemical physics},
  volume={107},
  number={23},
  pages={10278--10287},
  year={1997},
  publisher={AIP Publishing}
}

@article{doniach1970many,
  title={Many-electron singularity in X-ray photoemission and X-ray line spectra from metals},
  author={Doniach, Sunjic and Sunjic, Marijan},
  journal={Journal of Physics C: Solid State Physics},
  volume={3},
  number={2},
  pages={285--291},
  year={1970}
}

@article{mollenstedt1949electrostatic,
  title={The electrostatic lens as a high-resolution velocity analyzer},
  author={Möllenstedt, G},
  journal={Optics},
  volume={5},
  pages={499--517},
  year={1949}
}

@article{spencer2021inelastic,
  title={Inelastic background modelling applied to hard X-ray photoelectron spectroscopy of deeply buried layers: A comparison of synchrotron and lab-based (9.25 keV) measurements},
  author={Spencer, BF and Maniyarasu, Suresh and Reed, BP and Cant, DJH and Ahumada-Lazo, R and Thomas, AG and Muryn, CA and Maschek, M and Eriksson, SK and Wiell, T and others},
  journal={Applied Surface Science},
  volume={541},
  pages={148635},
  year={2021},
  publisher={Elsevier}
}

@article{van1979bulk,
  title={Bulk-and surface-plasmon-loss intensities in photoelectron, Auger, and electron-energy-loss spectra of Mg metal},
  author={Van Attekum, PM Th M and Trooster, JM},
  journal={Physical Review B},
  volume={20},
  number={6},
  pages={2335},
  year={1979},
  publisher={APS}
}

@article{seignac1972proprietes,
  title={Proprietes optiques de couches minces de platine dans l'ultra-violet lointain},
  author={Seignac, A and Robin, Mme S},
  journal={Solid State Communications},
  volume={11},
  number={1},
  pages={217--219},
  year={1972},
  publisher={Elsevier}
}

@article{gauthe1958contribution,
  title={Contribution {\`a} l’{\'e}tude des pertes d’{\'e}nergie subies par des {\'e}lectrons a la travers{\'e}e de couches m{\'e}talliques minces; comparaison avec les r{\'e}sultats de la spectroscopie des rayons X},
  author={Gauth{\'e}, Bernard},
journal={Annales de Physique},
  volume={13},
  pages={915--964},
  year={1958},
piblisher={EDP Sciences}
}

@article{kleinn1954energiespektren,
  title={Energiespektren von 35 kV Elektronen die an Festk{\"o}rperoberfl{\"a}chen reflektiert wurden},
  author={Kleinn, W},
  journal={Optik},
  volume={11},
  pages={226--243},
  year={1954}
}

@article{lynch1968characteristic,
  title={The characteristic loss spectra of the second and third series transition metals},
  author={Lynch, MJ and Swan, JB},
  journal={Australian Journal of Physics},
  volume={21},
  number={6},
  pages={811--816},
  year={1968},
  publisher={CSIRO Publishing}
}

@article{hufner1975core,
  title={Core-line asymmetries in the x-ray-photoemission spectra of metals},
  author={H{\"u}fner, S and Wertheim, GK},
  journal={Physical Review B},
  volume={11},
  number={2},
  pages={678},
  year={1975},
  publisher={APS}
}

@article{hufner1975xps,
  title={XPS core line asymmetries in metals},
  author={H{\"u}fner, S and Wertheim, GK and Wernick, JH},
  journal={Solid State Communications},
  volume={17},
  number={4},
  pages={417--422},
  year={1975},
  publisher={Elsevier}
}

@article{rincon2023platinum,
  title={Platinum thin film by Ag L$\alpha$, hard x-ray photoelectron spectroscopy},
  author={Rinc{\'o}n-Ortiz, Sergio A and Quintero-Orozco, Jorge H and Ospina, Rogelio},
  journal={Surface Science Spectra},
  volume={30},
  number={2},
  year={2023},
  publisher={AIP Publishing}
}

@article{tehuacanero2016low,
  title={The low-loss EELS spectra from radiation damaged gold nanoparticles},
  author={Tehuacanero-Cuapa, S and Reyes-Gasga, J and Rodr{\'\i}guez-G{\'o}mez, A and Bahena, D and Hern{\'a}ndez-Calder{\'o}n, I and Garc{\'\i}a-Garc{\'\i}a, R},
  journal={Journal of Applied Physics},
  volume={120},
  number={16},
  year={2016},
  publisher={AIP Publishing}
}

@article{leder1956comparison,
  title={Comparison of the characteristic energy losses of electrons with the fine structure of the x-ray absorption spectra},
  author={Leder, Lewis B and Mendlowitz, H and Marton, L},
  journal={Physical Review},
  volume={101},
  number={5},
  pages={1460},
  year={1956},
  publisher={APS}
}

@book{karlsson1967electron,
  title={Electron binding energies in platinum},
  author={Karlsson, SE and Norberg, CH and Nilsson, O and Hogberg, S and El-Farrash, AH and Nordling, C and Siegbahn, K},
  year={1967},
  publisher={Uppsala Univ., Inst. of Physics}
}

@article{nyholm1980core,
  title={Core level binding energies for the elements Hf to Bi (Z= 72-83)},
  author={Nyholm, Ralf and Berndtsson, Anders and Martensson, Nils},
  journal={Journal of Physics C: Solid State Physics},
  volume={13},
  number={36},
  pages={L1091},
  year={1980},
  publisher={IOP Publishing}
}

@article{shyu1988identification,
  title={Identification of platinum phases on $\gamma$-alumina by XPS},
  author={Shyu, JZ and Otto, K},
  journal={Applied Surface Science},
  volume={32},
  number={1-2},
  pages={246--252},
  year={1988},
  publisher={Elsevier}
}

@article{schon1972high,
  title={High resolution Auger electron spectroscopy of metallic copper},
  author={Sch{\"o}n, Gunnar},
  journal={Journal of Electron Spectroscopy and Related Phenomena},
  volume={1},
  number={4},
  pages={377--387},
  year={1972},
  publisher={Elsevier}
}

@article{schneider1981actinide,
  title={Actinide—Noble-metal systems: An x-ray-photoelectron-spectroscopy study of thorium-platinum, uranium-platinum, and uranium-gold intermetallics},
  author={Schneider, Wolf-Dieter and Laubschat, Clemens},
  journal={Physical Review B},
  volume={23},
  number={3},
  pages={997},
  year={1981},
  publisher={APS}
}

@article{lee2023phase,
  title={Phase Quantification of Heterogeneous Surfaces Using DFT-Simulated Valence Band Photoemission Spectra},
  author={Lee, Roxy and Quesada-Cabrera, Raul and Willis, Joe and Iqbal, Asif and Parkin, Ivan P and Scanlon, David O and Palgrave, Robert G},
  journal={ACS Applied Materials \& Interfaces},
  volume={15},
  number={33},
  pages={39956--39965},
  year={2023},
  publisher={ACS Publications}
}

@article{campbell2001widths,
  title={Widths of the atomic K--N7 levels},
  author={Campbell, JL and Papp, Tibor},
  journal={Atomic Data and Nuclear Data Tables},
  volume={77},
  number={1},
  pages={1--56},
  year={2001},
  publisher={Elsevier}
}

@techreport{perkins1991tables,
  title={Tables and graphs of atomic subshell and relaxation data derived from the LLNL Evaluated Atomic Data Library (EADL), Z= 1--100},
  author={Perkins, ST and Cullen, DE and Chen, MH and Rathkopf, J and Scofield, J and Hubbell, JH},
  year={1991},
  institution={Lawrence Livermore National Lab.(LLNL), Livermore, CA (United States)}
}

@article{fuggle1980core,
  title={Core-level lifetimes as determined by x-ray photoelectron spectroscopy measurements},
  author={Fuggle, John C and Alvarado, Santos F},
  journal={Physical Review A},
  volume={22},
  number={4},
  pages={1615},
  year={1980},
  publisher={APS}
}

@book{papaconstantopoulos1986handbook,
  title={Handbook of the Band Structure of Elemental Solids: From Z},
  author={Papaconstantopoulos, Dimitris A and others},
  year={1986},
  publisher={Springer}
}

@article{smidstrup2019quantumatk,
  title={QuantumATK: an integrated platform of electronic and atomic-scale modelling tools},
  author={Smidstrup, S{\o}ren and Markussen, Troels and Vancraeyveld, Pieter and Wellendorff, Jess and Schneider, Julian and Gunst, Tue and Verstichel, Brecht and Stradi, Daniele and Khomyakov, Petr A and Vej-Hansen, Ulrik G and others},
  journal={Journal of Physics: Condensed Matter},
  volume={32},
  number={1},
  pages={015901},
  year={2019},
  publisher={IOP Publishing}
}

@article{van2018pseudodojo,
  title={The PseudoDojo: Training and grading a 85 element optimized norm-conserving pseudopotential table},
  author={van Setten, Michiel J and Giantomassi, Matteo and Bousquet, Eric and Verstraete, Matthieu J and Hamann, Don R and Gonze, Xavier and Rignanese, G-M},
  journal={Computer Physics Communications},
  volume={226},
  pages={39--54},
  year={2018},
  publisher={Elsevier}
}

@article{perdew2008PBEsol,
  title={Restoring the density-gradient expansion for exchange in solids and surfaces},
  author={Perdew, John P and Ruzsinszky, Adrienn and Csonka, G{\'a}bor I and Vydrov, Oleg A and Scuseria, Gustavo E and Constantin, Lucian A and Zhou, Xiaolan and Burke, Kieron},
  journal={Physical Review Letters},
  volume={100},
  number={13},
  pages={136406},
  year={2008},
  publisher={APS}
}
\bibliographystyle{apsrev4-1}

\end{document}


\preprint{APS/123-QED}

\title[PRB]{Lifetime effects and satellites in the photoelectron spectrum of platinum metal}

\newcommand{\OxfordChem}{Department of Chemistry, University of Oxford, Inorganic Chemistry Laboratory, South Parks Road, Oxford, OX1 3QR, United Kingdom}
\newcommand{\UCLChem}{Department of Chemistry, University College London, 20 Gordon Street, London, WC1H 0AJ, United Kingdom}

\author{P.~Bhatt}
\affiliation{\protect\UCLChem}
\affiliation{\protect\OxfordChem}
\affiliation{Istituto Officina dei Materiali (IOM)-CNR, Laboratorio TASC, in Area Science Park, S.S.14, Km 163.5, Trieste I-34149, Italy}

\author{J.~J.~Gutiérrez Moreno}
\affiliation{Barcelona Supercomputing Center, Plaça Eusebi Güell 1-3, 08034 Barcelona, Spain}

\author{L.~E.~Ratcliff}
\affiliation{Centre for Computational Chemistry, School of Chemistry, University of Bristol, Bristol BS8 1TS, United Kingdom}
\affiliation{Hylleraas Centre for Quantum Molecular Sciences, Department of Chemistry, UiT The Arctic University of Norway, N-9037 Tromsø, Norway}

\author{A.~A.~Riaz}
\affiliation{\protect\UCLChem}

\author{C.~M.~L.~André}
\affiliation{\protect\OxfordChem}
\affiliation{Université Paris-Saclay, 9 Rue Joliot Curie, Gif-sur-Yvette, 91190 France}

\author{A.~S.~Y.~Lu}
\affiliation{\protect\OxfordChem}

\author{R.~G.~Palgrave}
\affiliation{\protect\UCLChem}
\affiliation{HarwellXPS, Research Complex at Harwell, Rutherford Appleton Laboratory, Didcot OX11 0DE, United Kingdom}

\author{A.~Gloskovskii}
\author{C.~Schlueter}
\affiliation{Deutsches Elektronen-Synchrotron (DESY), Notkestraße 85, Hamburg 22607, Germany}

\author{P.~K.~Thakur}
\author{T.-L.~Lee}
\affiliation{Diamond Light Source Ltd., Diamond House, Harwell Science and Innovation Campus, Didcot, OX11 0DE, United Kingdom}

\author{A.~Regoutz}
\affiliation{\protect\UCLChem}
\affiliation{\protect\OxfordChem}

 \email{anna.regoutz@chem.ox.ac.uk}

\date{\today}
\maketitle


 \tableofcontents

\cleardoublepage

\section{Additional Experimental Details}
REELs on Pt were performed at the UK National XPS Facility, HarwellXPS. All experiments were performed on a Thermo Scientific\textsuperscript{TM} NEXSA G2 spectrometer. The instrument employs a monochromated X-ray Al K$\alpha$ source (h = 1486.6 eV), Helium I and II UV sources, an electron source flood gun, an $180^{\circ}$ hemispherical analyser, and a two–dimensional detector. To provide a clean surface on the Pt foil, the measured spots were first sputtered \textit{in-situ} using a focused Ar\textsuperscript{+} ion gun operating at 2~keV until O and C signals were minimised. REELS measurements were conducted using the electron flood gun as the electron source, with beam energies of 0.25, 0.5, 0.75, and 1~keV and an emission current of 5~$\mu$A. A pass energy of 40~eV was used to collect the REELS data. The experimental resolution across all energies taken from the full width half maximum of the main elastic peak is 517~meV. \par

\section{Energy Resolution Determination and Parameters}

\begin{figure*}[ht!]
\centering
    \includegraphics[keepaspectratio, width = 0.95\linewidth]{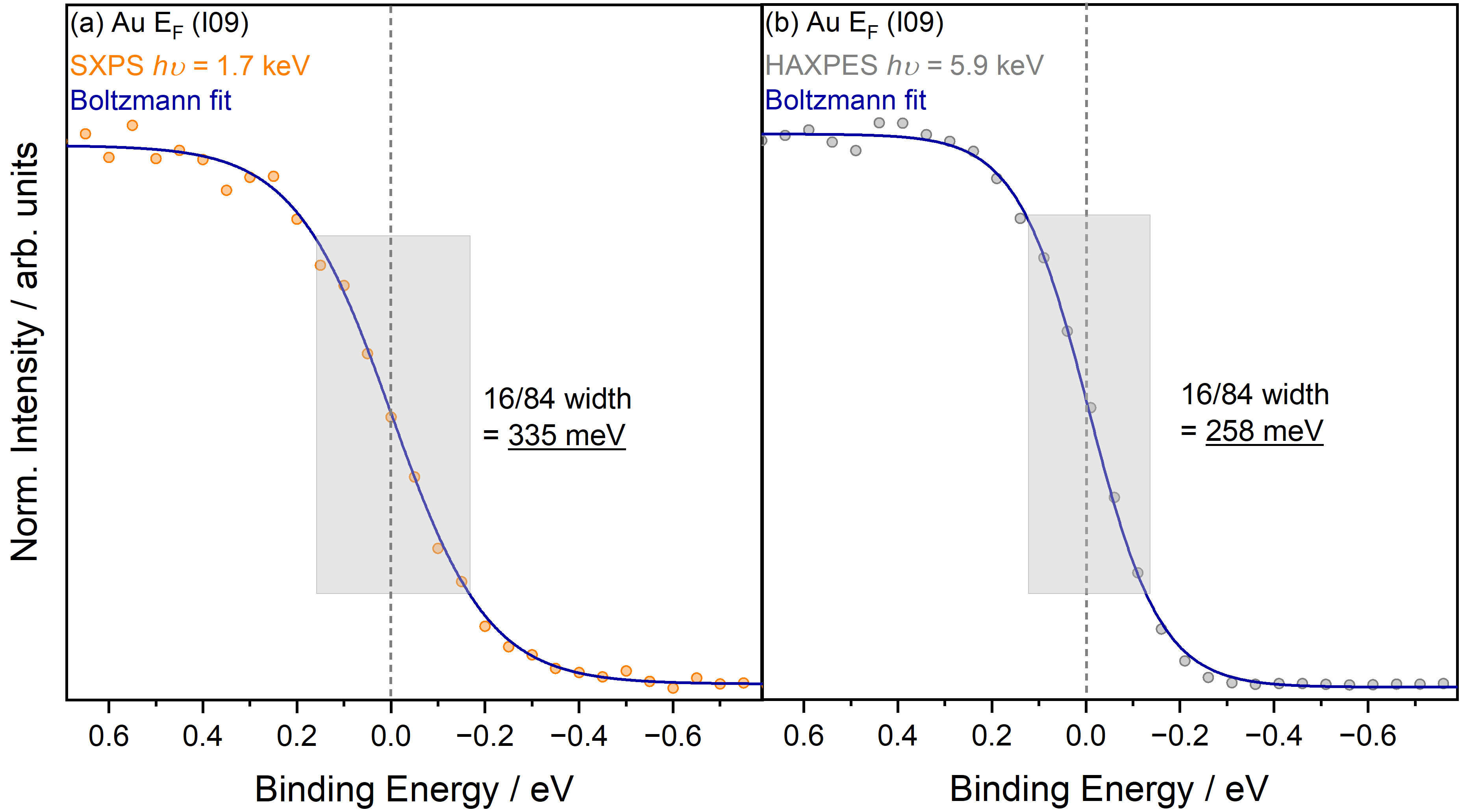}
    \caption{Fermi edges (E\textsubscript{F}) of a high-purity gold (Au) polycrystalline reference foil, collected at beamline I09, Diamond Light Source. The E\textsubscript{F} was collected with photon energies labelled as (b) SXPS ($h\nu$ = 1.7~keV) and (b) HAXPES ($h\nu$ = 5.9~keV) energies, described in the main manuscript. The total experimental resolution was determined using the 16/84\% method, as shown in the grey box. The data were fit with a Boltzmann curve to identify the E\textsubscript{F} energy position.}
    \label{fig:Au}
\end{figure*}

\section{QUASES-IMFP-TPP2M Input Parameters}

The input parameters for Pt for inelastic mean free path (IMFP) calculations using the QUASES-IMFP-TPP2M~v3.0 software package are summarised in Table~\ref{table:Quases}.\par

\begin{table}[hb]
    \centering
       \caption{Input parameters from QUASES-IMFP-TPP2M~v3.0 for Pt for the determination of the inelastic mean free path (IMFP). The IMFP is calculated using the non-relativistic TPP2M formula.~\cite{tanuma1994calculations}}
    \begin{tabular}{lc}
    \hline \hline
        QUASES input parameter & value \\
        \hline 
        Density / gcm\textsuperscript{-3} & 21.4 \\
        No. of s+p(lbe) valence electrons per atom & 1 \\
        No. of d valence electrons per atom & 9 \\
        atomic mass / $u$ & 195.09 \\
        Band gap energy / eV & 0 \\
            \hline \hline
    \end{tabular}

    \label{table:Quases}
\end{table}

\cleardoublepage

\section{G\textsubscript{0}W\textsubscript{0} and DFT Calculated Unweighted PDOS}
\begin{figure}[hb!]
\centering
    \includegraphics[keepaspectratio, width = 0.75\linewidth]{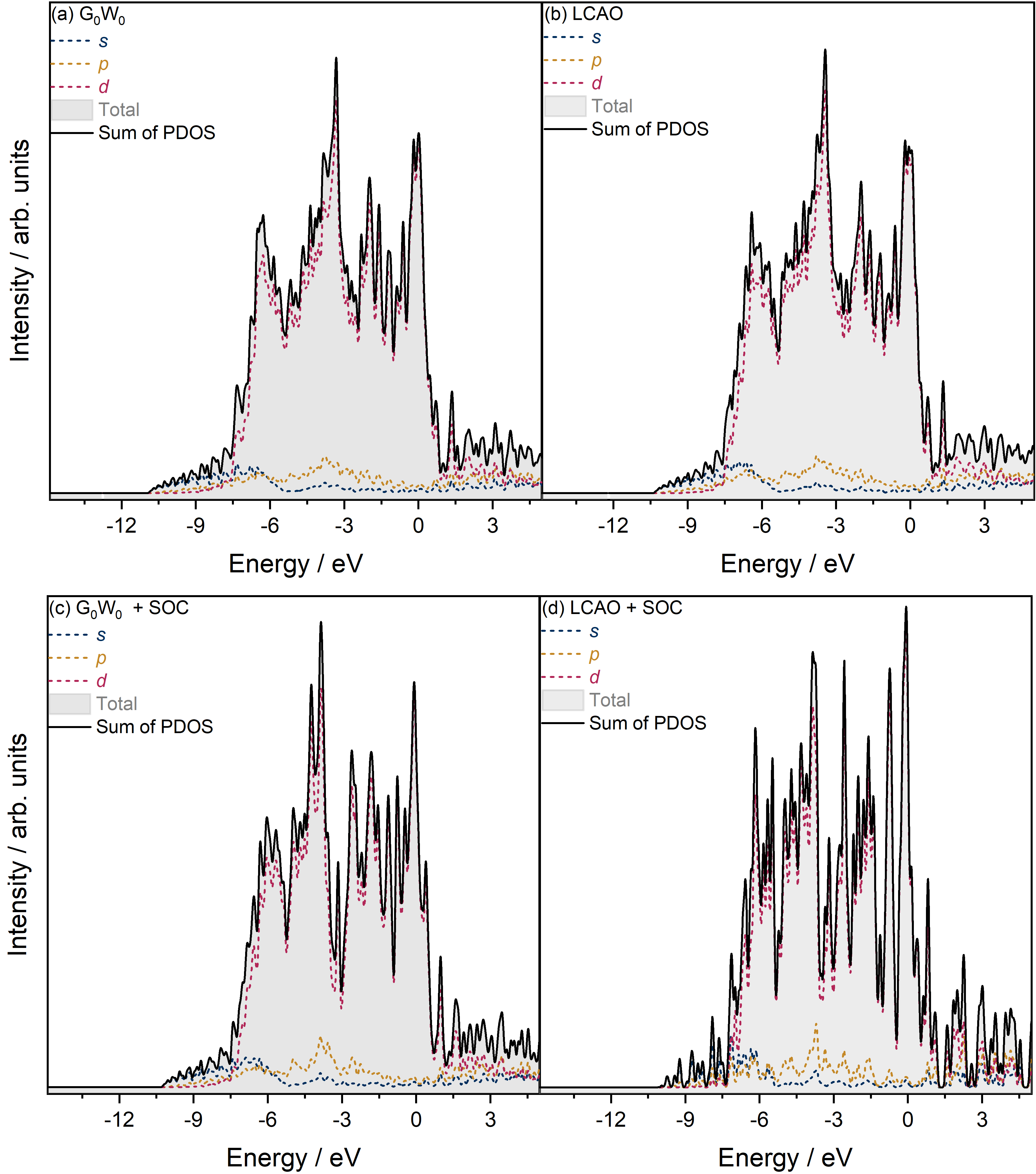}
    \caption{Unweighted, projected DOS (PDOS) calculated using the (a,c) G\textsubscript{0}W\textsubscript{0}, and (b,d) DFT approaches. Subfigures (c,d) are calculated with consideration of SOC. The PDOS contributions are represented by dashed lines. The black line represents the sum of the PDOS, while the grey area labelled `total' corresponds to the total DOS obtained directly from the theoretical calculations.}
    \label{fig:DOS}
\end{figure}

Figure~\ref{fig:DOS} shows the unweighted DOS spectra calculated using the two computational approaches, (a,c) G\textsubscript{0}W\textsubscript{0}, and (b,d) DFT with and without the inclusion of spin-orbit coupling (SOC). No obvious differences are observed between the sum of PDOS calculated manually and the total DOS provided by the theoretical calculations, suggesting complete projection of states. 

\cleardoublepage

\section{Additional Information on R(H)EELS Experiments and Summary of Literature}
Table~\ref{tab:reelslit} summarises all known reports of features from electron energy loss spectroscopic investigations. All features are tabulated with $\omega$ positions, relative to the main elastic peak. Where investigations have discussed the origin of the features, the assignment has been given otherwise, the phrase `not provided' is stated.

\begin{table}[ht!]
\caption{\label{tab:reelslit} Energy loss ($\omega$) peak positions and assignments of loss features of Pt metal from literature determined from REELS. Where `Not provided' is written, the work only reports the positions of features without interpretation of their origin.}

    \begin{tabular}{ccc}
    \hline \hline
    $\omega$ of Feature / eV& Origin of Feature& Reference \\
    \hline
6.6 & \multirow{5}{*}{Not provided} & \multirow{5}{*}{\cite{rudberg1930characteristic}} \\
9.4 & &  \\
11.7 & &  \\
33.7 & &  \\
34.8& &  \\
\hline    
14 & \multirow{3}{*}{Not provided} & \multirow{5}{*}{\cite{mollenstedt1949electrostatic},\cite{albert1956einfluss}} \\
22.4 & &  \\
46 & & \\
47.3 & Plasmonic overtone& \\
61.4& Not provided& \\
\hline
18.5 & \multirow{5}{*}{Not provided} &\multirow{5}{*}{\cite{gauthe1958contribution}} \\
23.9 & & \\
30 & & \\
37 & & \\
47& & \\
\hline
5.2 & \multirow{2}{*}{Not provided} & \multirow{2}{*}{\cite{kleinn1954energiespektren}} \\
22.6& & \\
\hline
6.2 & \multirow{3}{*}{Not provided}& \multirow{3}{*}{\cite{powell1960origin}}\\
14.3 & & \\
22.4& & \\
\hline \hline
\end{tabular}
\quad
\begin{tabular}{ccc}
    \hline \hline
    $\omega$ of Feature / eV& Origin of Feature& Reference \\
    \hline
3.6 & \multirow{5}{*}{Interband transitions} &\multirow{7}{*}{\cite{seignac1972proprietes}} \\
6.3& &\\  
9.8& &\\  
14.5& &\\ 
19.8& &\\  
24& Surface plasmon&\\  
33.5& Bulk plasmon&\\ 
\hline
7.8& Interband transitions&\multirow{4}{*}{\cite{lynch1968characteristic}} \\  
28.2& Bulk plasmon&\\  
55.2& Ionisation of the O\textsubscript{III}-level&\\  
73.4& ‘G’ peaks from excited state transitions&\\  
\hline
7.4& Interband transitions&\multirow{7}{*}{\cite{schroder1974investigation}} \\  
13.5& Not provided&\\  
24.8& Surface plasmon&\\  
31.8& Bulk plasmon&\\  
45.1& Not provided&\\  
54.1& Ionisation of the O\textsubscript{III}-level&\\  
71.2& 4\textit{f} ionisation& \\
\hline
\multirow{2}{*}{47.3}& \multirow{2}{*}{Plasmonic overtone}& \multirow{2}{*}{\cite{leder1956comparison}}\\
& & \\
\hline \hline   
\end{tabular}
\end{table}

\begin{figure}[htbp]
\centering
    \includegraphics[keepaspectratio, width = 0.7\linewidth]{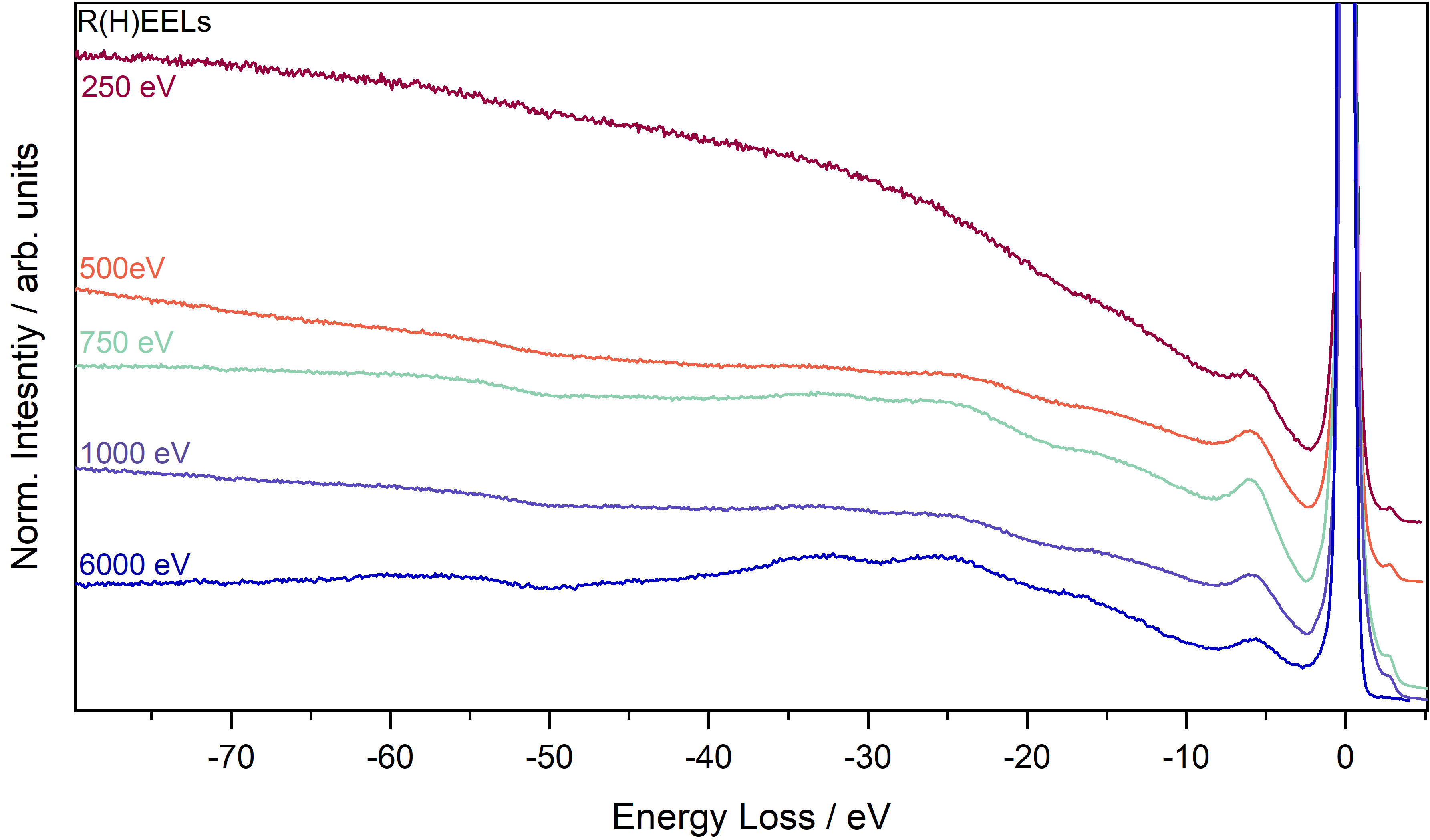}
    \caption{R(H)EELS spectra of metallic platinum measured at different electron energies, including 250~eV, 500~eV, 750~eV, 1000~eV, and 6000~eV. Spectra were aligned to the primary elastic peak and normalised to its height.}
    \label{fig:REELsoverview}
\end{figure}

\begin{figure}[htbp]
\centering
    \includegraphics[keepaspectratio, width = 0.5\linewidth]{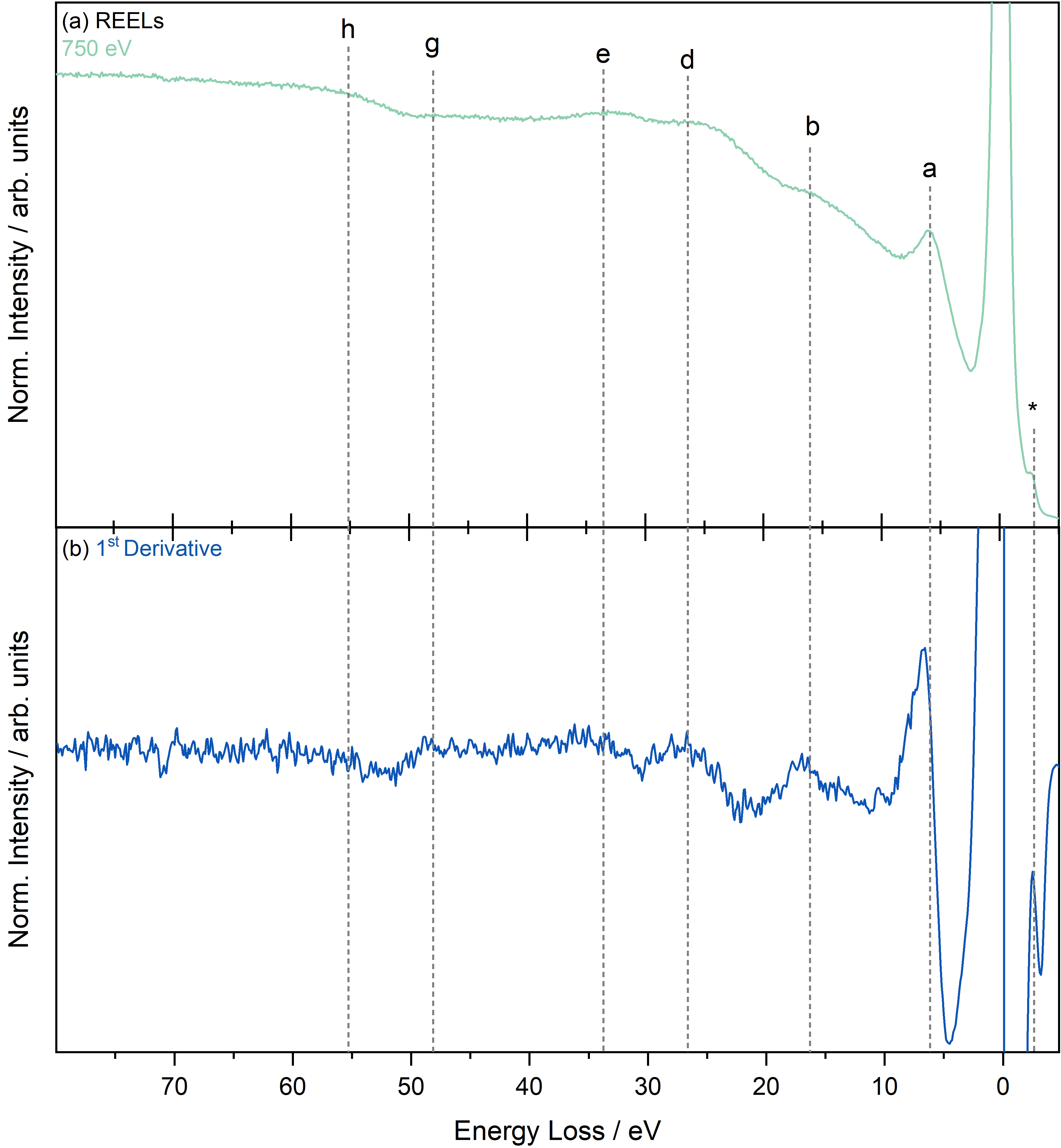}
    \caption{REELS spectrum of metallic platinum collected at an electron energy of 750~eV. (a) Collected REELS spectrum, and (b) its first derivative with respect to energy loss. The energy-loss features are labelled \textbf{a-h}. The spectra are aligned to the primary elastic peak. The first-derivative curve was smoothed using a 2\textsuperscript{nd}- order Savitzky-Golay method over a window of 22 data points.}
    \label{fig:REELS}
\end{figure}

 Figure~\ref{fig:REELsoverview} shows the R(H)EELs spectra collected at various electron energies. The loss features are observed after the main elastic peak at higher energy-loss positions. The RHEELs spectrum collected at 6000~eV in blue shows the greatest number of resolved features, primarily due to the improved energy resolution of the detector and lower relative background, and is thus the main focus of the discussion of the electronic structure of Pt in the main manuscript. However, the observed features of both methods agree well (see Figure~\ref{fig:REELS} for the 750~eV data, including first derivative).
 
 \begin{figure}[htbp]
\centering
    \includegraphics[keepaspectratio, width = 0.5\linewidth]{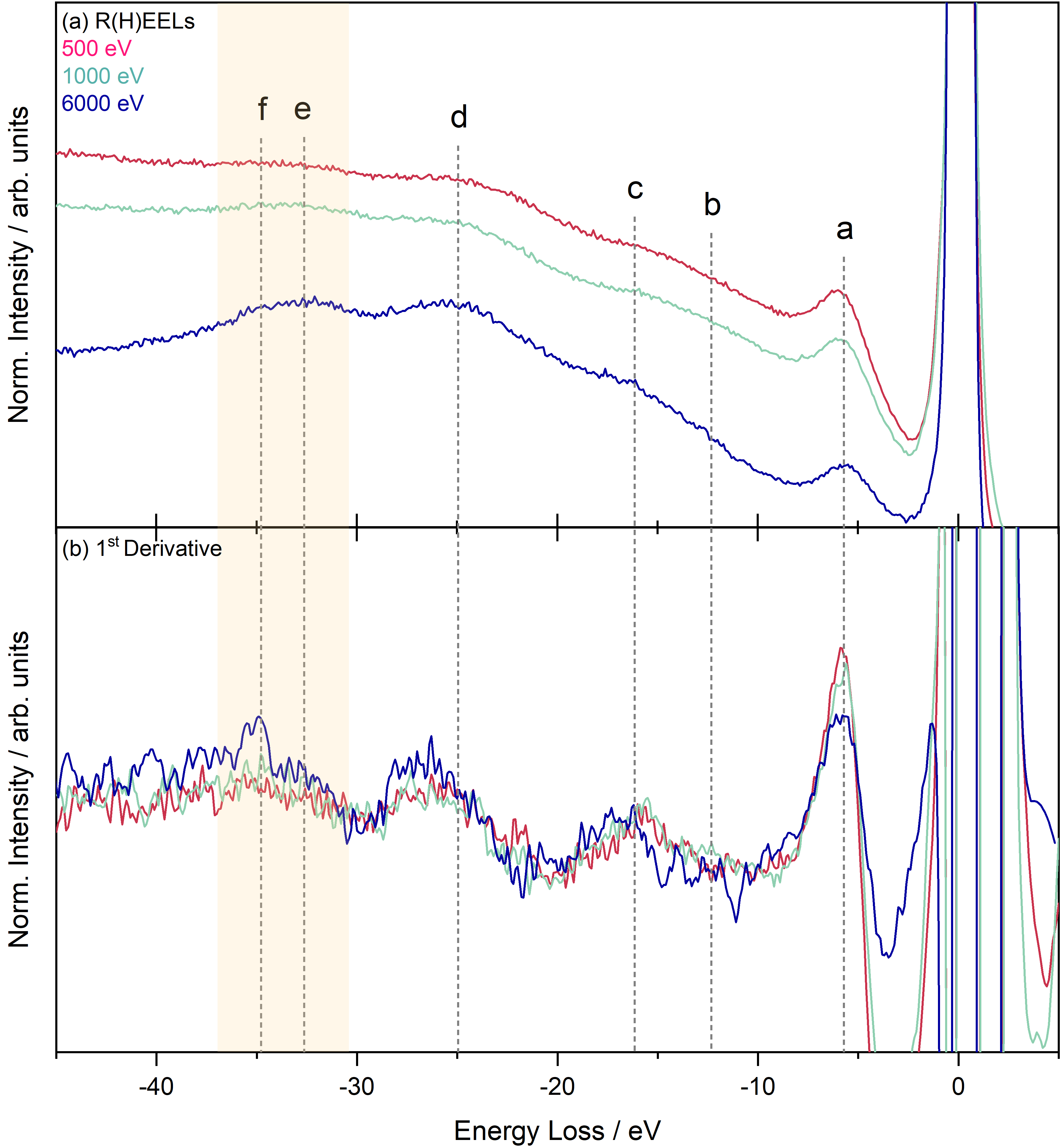}
    \caption{Zoomed in R(H)EELS spectrum of metallic platinum collected at different electron energies, including 500~eV, 1000~eV, and 6000~eV, showing the (a) Collected spectrum, and (b) its first derivative with respect to energy loss. The energy-loss features are labelled \textbf{a-f}. The spectra are aligned to the primary elastic peak. The first-derivative curve was smoothed using a 2\textsuperscript{nd}- order Savitzky-Golay method over a window of 22 data points. The boxed region containing features \textbf{e} and \textbf{f} are described in the main text.}
    \label{fig:REELSzoom}
\end{figure}

\cleardoublepage
\section{Survey Spectra}

\begin{figure}[ht!]
\centering
    \includegraphics[keepaspectratio, width = 0.75\linewidth]{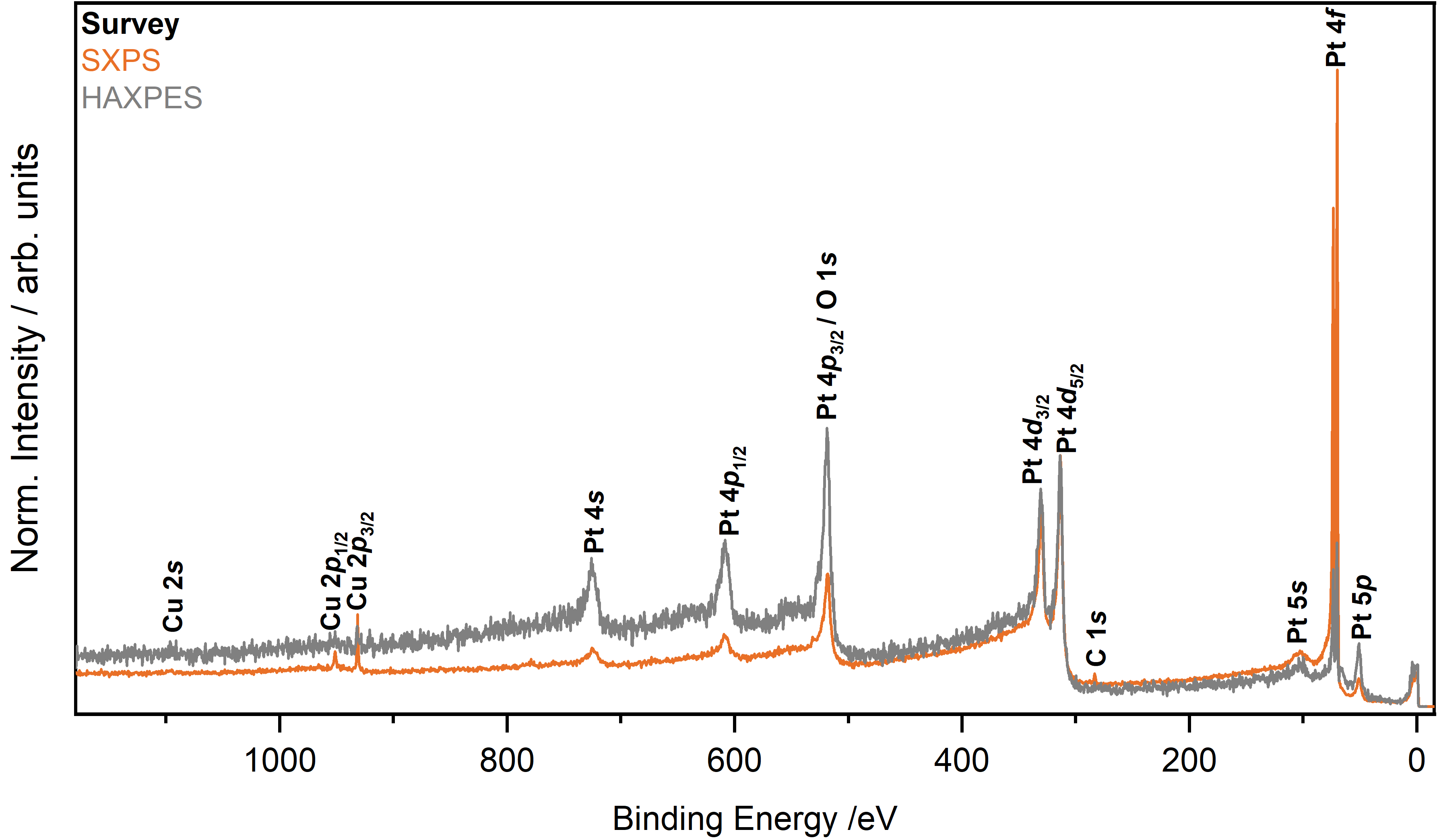}
    \caption{Survey spectra collected with synchrotron-based SXPS ($h\nu$ = 1.7~keV) and HAXPES ($h\nu$ = 5.9~keV) on a polycrystalline Pt foil. The BE is aligned to the intrinsic $E_F$ of the Pt metal, and the spectra are normalised to the height of the Pt~4\textit{d}\textsubscript{5/2} core level peak after linear background removal.}
    \label{fig:Survey}
\end{figure}

\begin{figure}[hb!]
\centering
    \includegraphics[keepaspectratio, width = 0.75\linewidth]{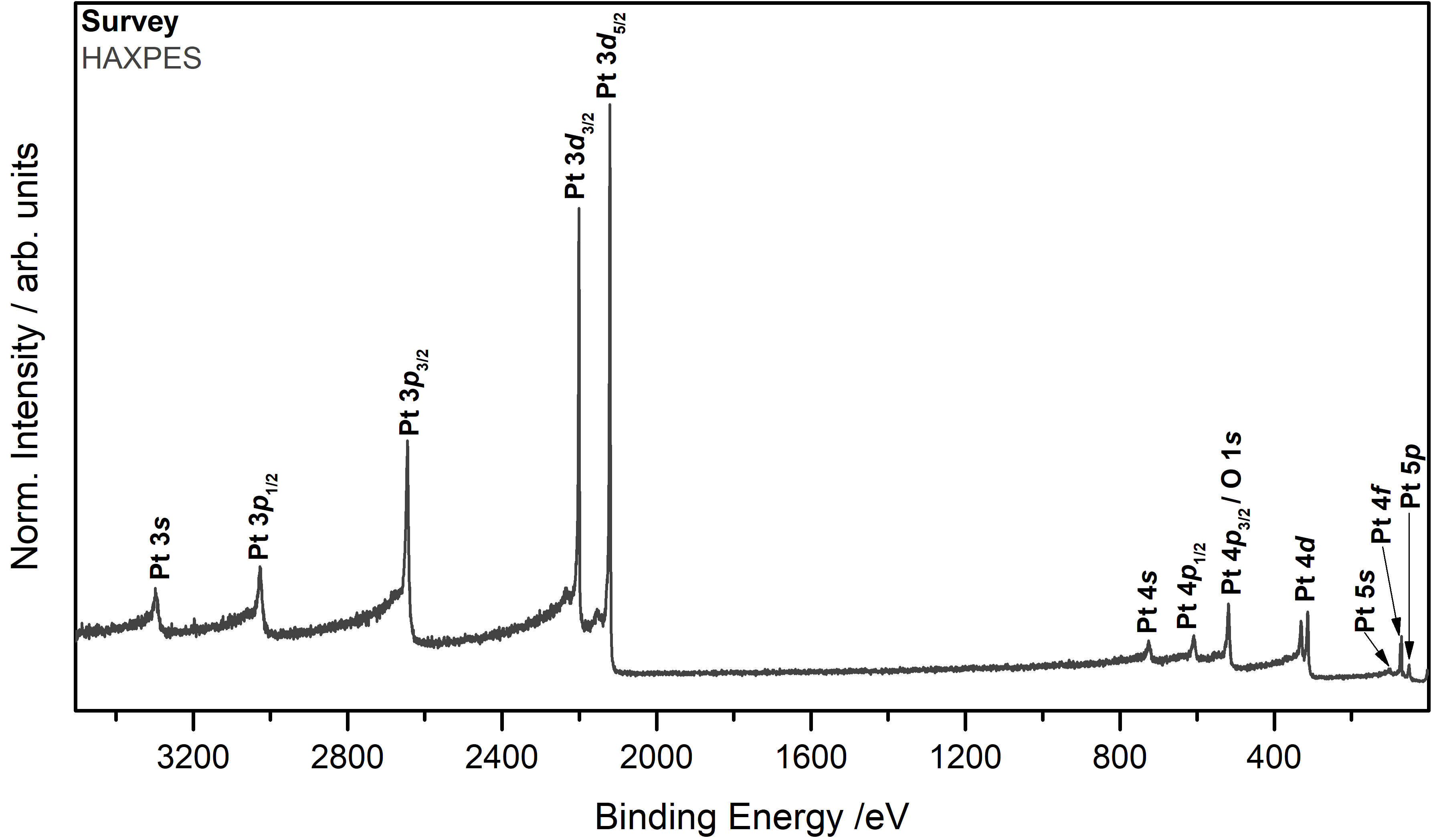}
    \caption{An extended BE range survey spectrum collected with synchrotron-based HAXPES ($h\nu$ = 5.9~keV) on a polycrystalline Pt foil. The BE is aligned to the intrinsic $E_F$ of the Pt metal.}
    \label{fig:Survey_hx}
\end{figure}

\cleardoublepage
\section{Further Photoionisation Cross Section Information}
\begin{figure}[!hb]
\centering
    \includegraphics[keepaspectratio, width = 0.5\linewidth]{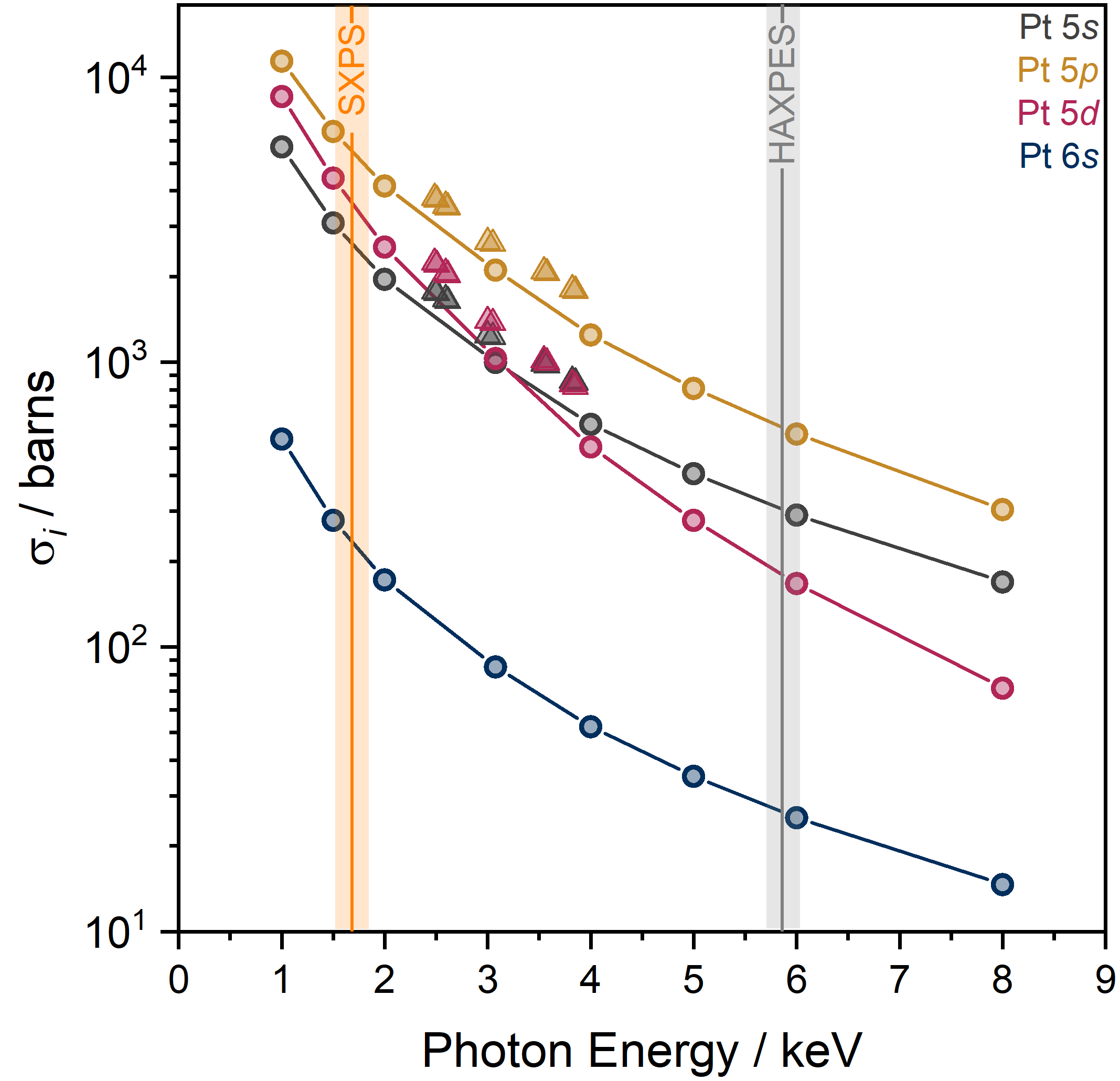}
    \caption{The one-electron corrected photoionisation cross sections $\sigma_i$ of relevant atomic orbitals at different photon energies. The maximum kinetic energies used for SXPS ($h\nu = 1.7$~keV) and HAXPES ($h\nu = 5.9$~keV) experiments are highlighted with tie lines. $\sigma_i$ values that are affected by the X-ray absorption of PT have been plotted as triangular scatter points.}
    \label{fig:cs}
\end{figure}

Figure~\ref{fig:cs} displays the one-electron photoionisation cross sections ($\sigma_i$) of key atomic orbitals of platinum (Pt) taken from Refs.~\cite{Scofield1973,Dig_Sco_2020} as a function of photon energy.

For the SXPS and HAXPES experiments conducted, Table~\ref{tab:cs} lists the $\sigma_i$ values of interest from Pt and Pb. These were used for the `Pb correction' approach for estimation of Pt~6\textit{p}, as described in the main manuscript. The correction factor determined for the PDOS is also shown in the Table. The `Pb correction' cross section values are determined by multiplying the Pt~6\textit{s} cross section by the Pb~6\textit{p}/Pb~6\textit{s} ratio, done in a previous work on W metal.~\cite{kalha2022lifetime}

\begin{table}[ht!]
\caption{\label{tab:cs}Tabulated one-electron photoionisation cross sections $\sigma_i$ of interest for Pt and Pb taken from Refs.\cite{Scofield1973,Dig_Sco_2020}. Values are shown for photon energies used in SXPS ($h\nu = 1.7$~keV) and HAXPES ($h\nu = 5.9$~keV) experiments. The correction factor from the ratio of the Pb Pb~6\textit{p}/Pb~6\textit{s} $\sigma_i$ values are also shown.}
    \begin{ruledtabular}
    \begin{tabular}{ccc}
    \textbf{Orbital} & \textbf{$\sigma_i$ (SXPS) / barns} & \textbf{$\sigma_i$ (HAXPES) / barns}\\
    \hline
    Pt~5\textit{p} & 2.85e+3& 2.90e+2\\
    Pt~5\textit{d}& 1.76e+3& 8.78e+1\\
    Pt~6\textit{s} & 2.26e+2& 2.58e+1\\
    Pb~6\textit{s} & 4.11e+2& 4.74e+1\\
    Pb~6\textit{p} & 2.42e+2& 2.83+1\\
    \hline
    Pb~6\textit{p}/Pb~6\textit{s} ratio & 5.89e-1& 5.97e-1\\
    Pb correction & 1.33e+2& 1.54e+1\\
    \end{tabular}
    \end{ruledtabular}
\end{table}
\cleardoublepage
\section{Comparison of Core Level Satellites}

\begin{figure}[ht!]
\centering
    \includegraphics[keepaspectratio, width = 0.65\linewidth]{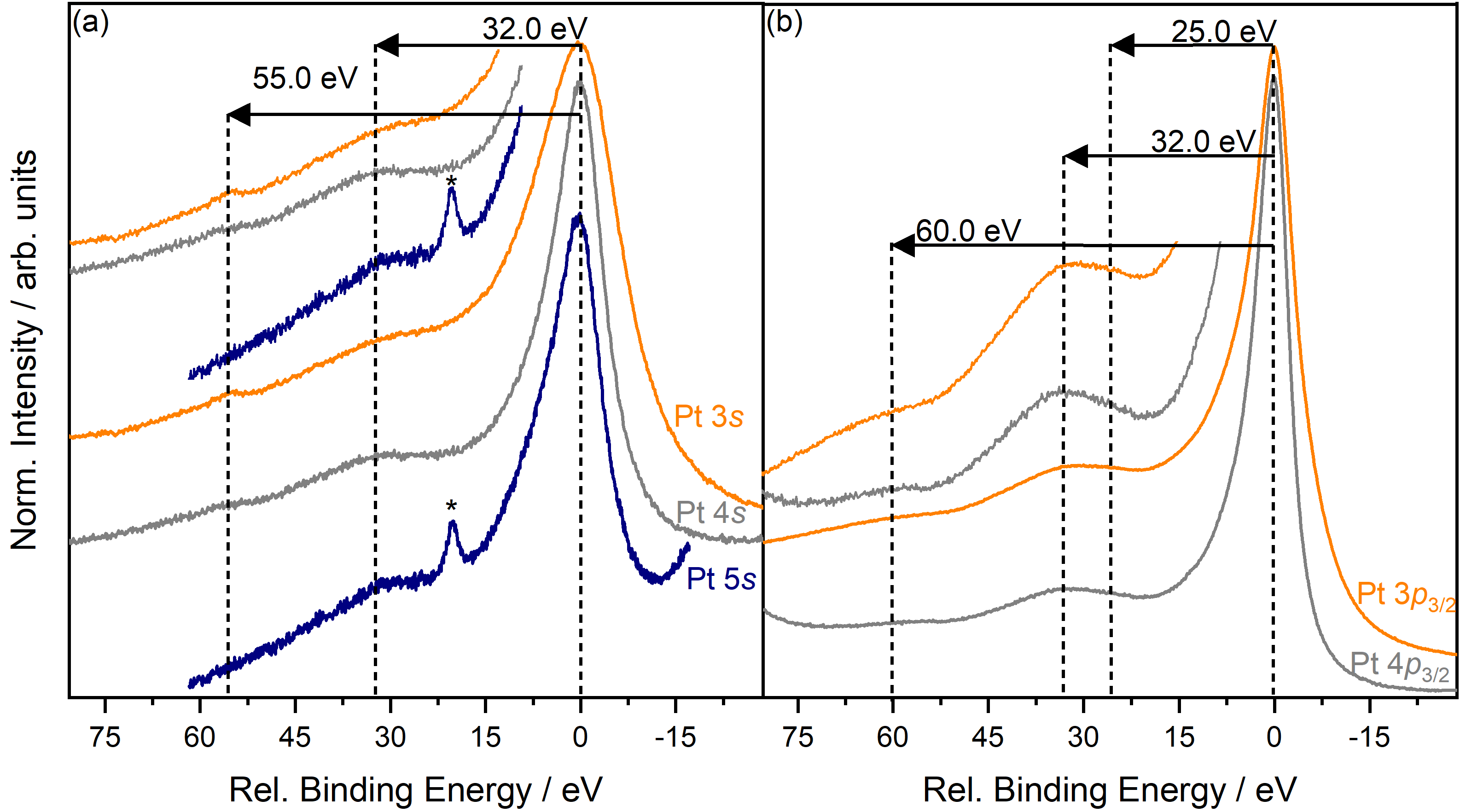}
    \caption{Comparison of the (a) Pt~3\textit{s}, Pt~4\textit{s} and (b) Pt~3\textit{p}\textsubscript{3/2}, Pt~4\textit{p}\textsubscript{3/2} core level HAXPES spectra. Spectra are normalised to their maximum intensity and aligned relative to the main photoionisation peak. Insets show zoomed-in regions of the higher binding energy tails.}
    \label{fig:ptspcomp}
\end{figure}

\section{Comparison of PDOS to XPS Valence Band}
\begin{figure*}[htbp]
\centering
    \includegraphics[keepaspectratio, width =0.9 \linewidth]{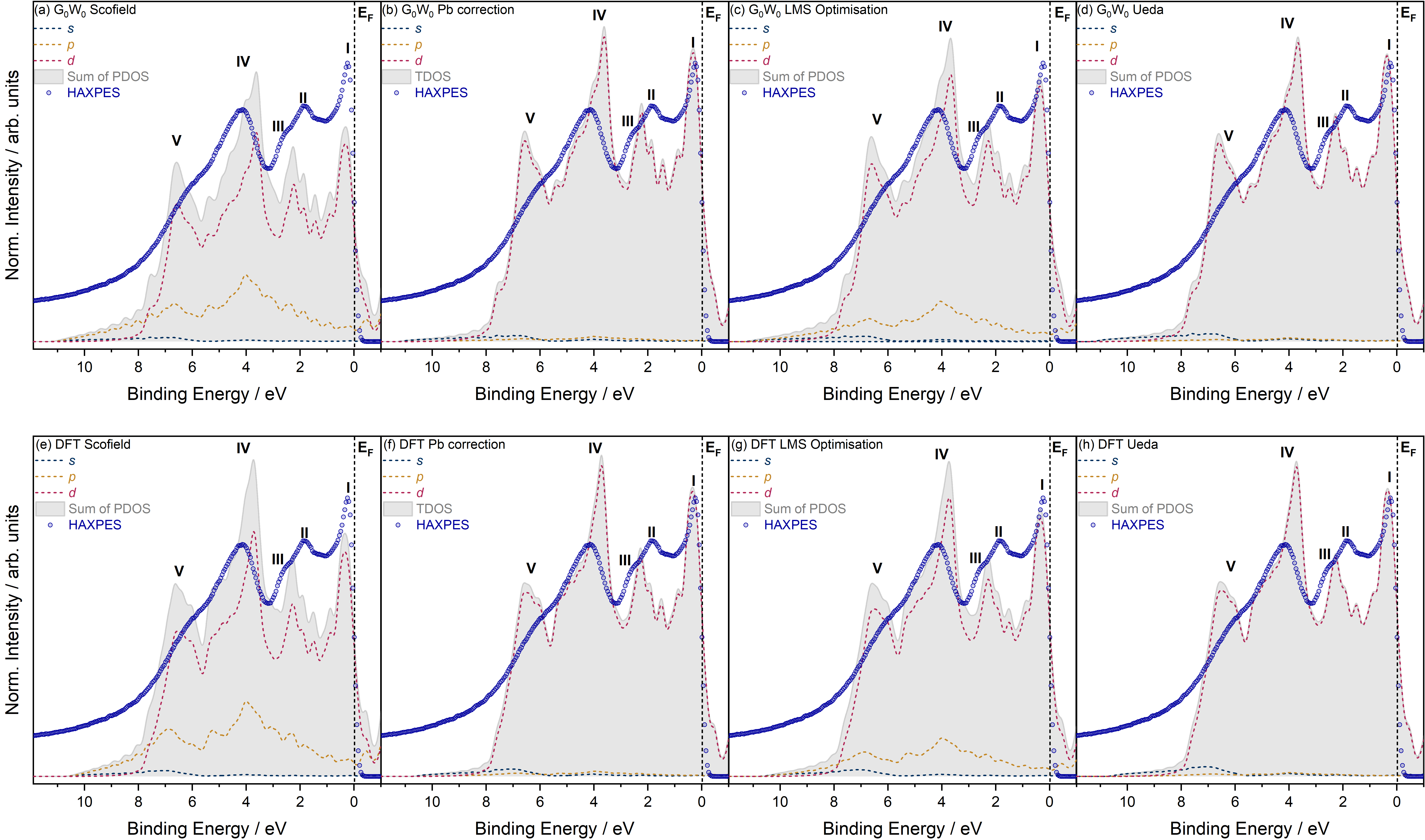}
    \caption{Comparison of the PDOS spectra calculated using G\textsubscript{0}W\textsubscript{0} (a-d) DFT (e-h) with HAXPES valence band spectra, including the (a,e) Scofield, (b,f) Pb correction, (c,g) LMS optimisation, and (d,h) Ueda \textit{p}-state cross section weighting approaches. The PDOS contributions have been broadened to match the experimental broadening. HAXPES spectra are normalised to their respective areas after the removal of a Shirley-type background.}
    \label{fig:HX_SI}
\end{figure*}
\cleardoublepage
\begin{figure}[htbp]
\centering
    \includegraphics[keepaspectratio, width = 0.49\linewidth]{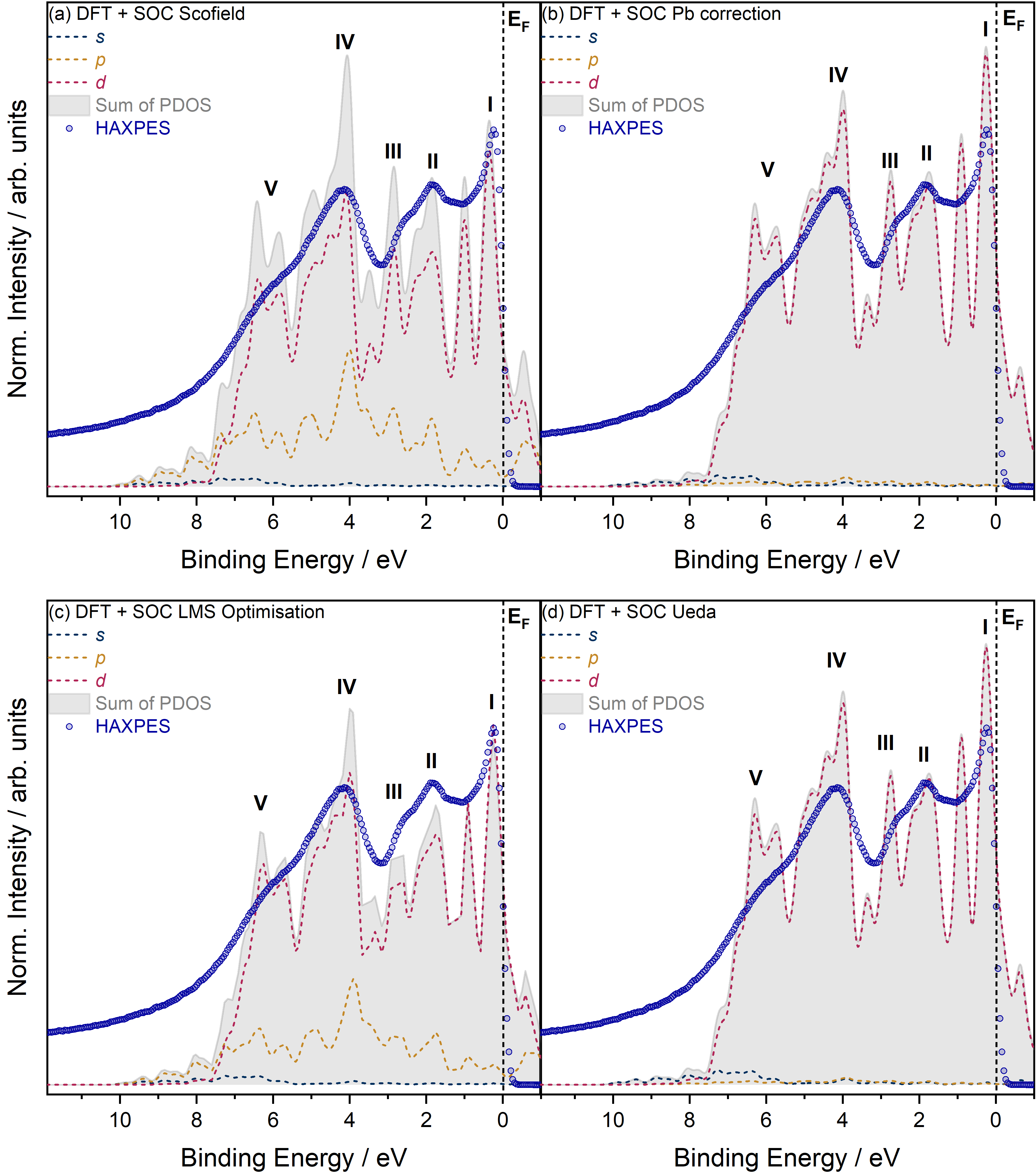}
    \caption{Comparison of the PDOS spectra calculated using DFT with SOC with HAXPES valence band spectra, including the (a) Scofield, (b) Pb correction, (c) LMS optimisation, and (d) Ueda \textit{p}-state cross section weighting approaches. The PDOS contributions have been broadened to match the experimental broadening. HAXPES spectra are normalised to their respective areas after the removal of a Shirley-type background.}
    \label{fig:HX_SOC_final}
\end{figure}

\begin{figure*}[!hb]
\centering
    \includegraphics[keepaspectratio, width = 0.64\linewidth]{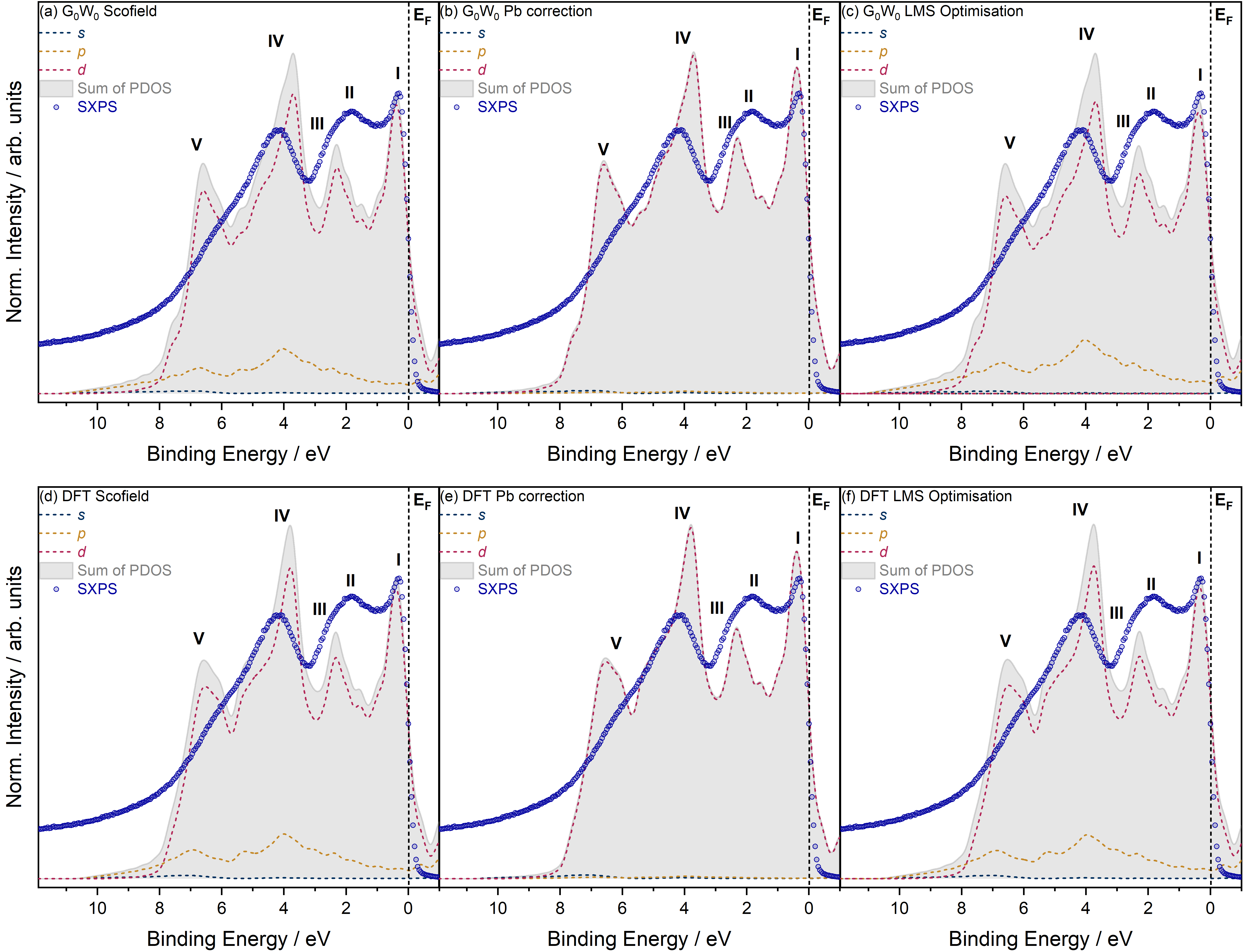}
    \caption{Comparison of the PDOS spectra calculated using G\textsubscript{0}W\textsubscript{0} (a-c) DFT (d-f) with SXPS valence band spectra, including the (a,d) Scofield, (b,e) Pb correction, and (c,f) LMS optimisation \textit{p}-cross section weighting approaches. The PDOS contributions have been broadened to match the experimental broadening. SXPS spectra are normalised to their respective areas after the removal of a Shirley-type background.}
    \label{fig:SX_final}
\end{figure*}

\clearpage

\begin{figure*}[htbp]
\centering
    \includegraphics[keepaspectratio, width = 0.64\linewidth]{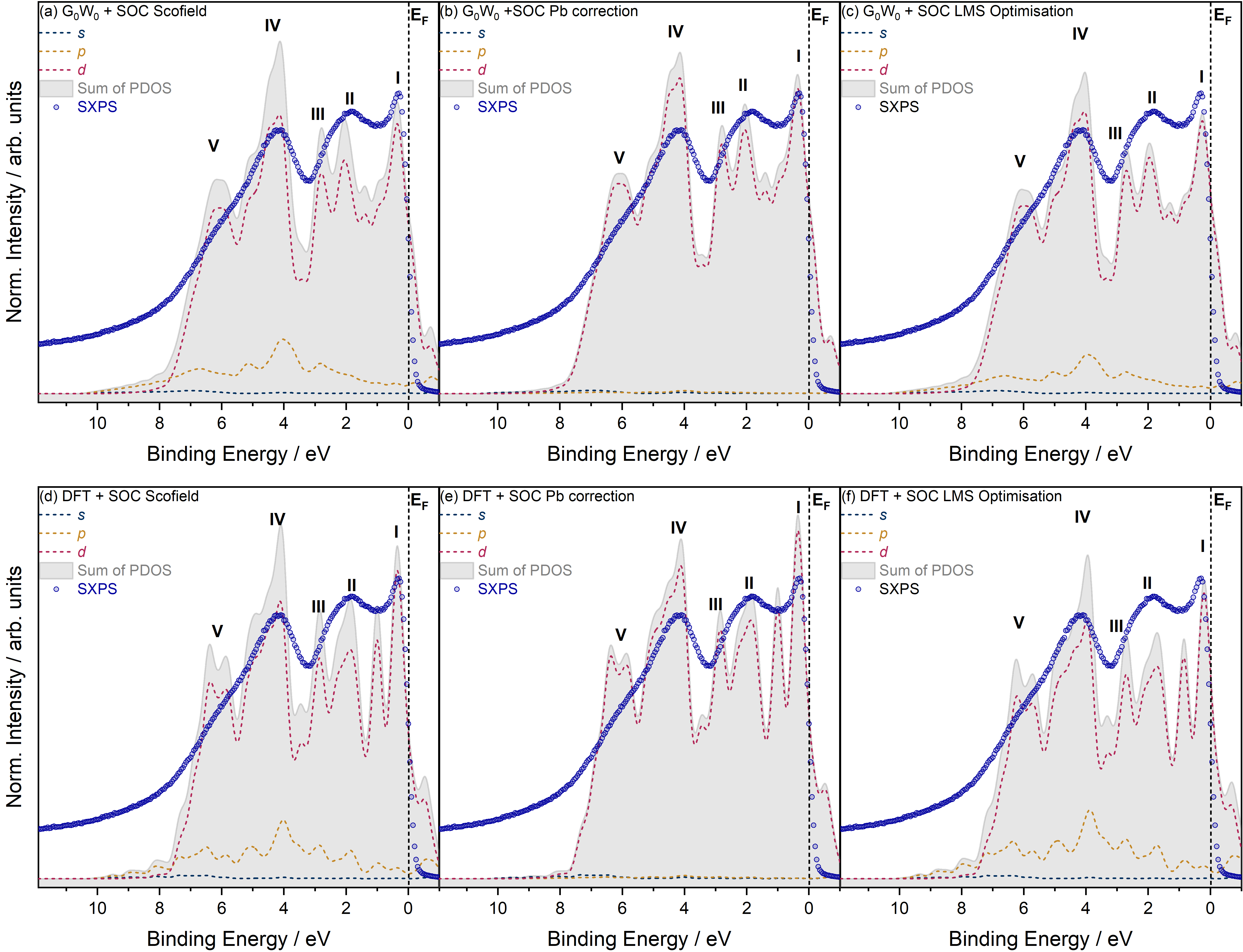}
    \caption{Comparison of the PDOS spectra calculated using G\textsubscript{0}W\textsubscript{0} (a-c) DFT (d-f) with SOC with SXPS valence band spectra, including the (a,d) Scofield, (b,e) Pb correction, and (c,f) LMS optimisation \textit{p}-state cross section weighting approaches. The PDOS contributions have been broadened to match the experimental broadening. SXPS spectra are normalised to their respective areas after the removal of a Shirley-type background.}
    \label{fig:SX_SOC_final}
\end{figure*}

\bibliography{references.bib}
\bibliographystyle{apsrev4-1}